\documentclass[aps,prd,showpacs,superscriptaddress,preprintnumbers]{revtex4}
\usepackage{graphicx,float,wrapfig,subfigure}
\usepackage{amsfonts,amsmath,amssymb,amstext}
\usepackage{latexsym}
 \usepackage{bm}
\usepackage{color}
\usepackage[normalem]{ulem}

\newcommand{\be}{\begin{equation}}
\newcommand{\ee}{\end{equation}}
\newcommand{\ba}{\begin{eqnarray}}
\newcommand{\ea}{\end{eqnarray}}

\definecolor{red}{rgb}{0.7,0,0}
\definecolor{green}{rgb}{0,0.5,0}

\begin{document}

\title{Rate and ellipticity of dilepton production in a magnetized quark-gluon plasma}
\date{July 14, 2022}

\author{Xinyang Wang}
\email{wangxy@ujs.edu.cn}
\affiliation{Department of Physics, Jiangsu University, Zhenjiang 212013, People’s Republic of China}

\author{Igor A. Shovkovy}
\email{igor.shovkovy@asu.edu}
\affiliation{College of Integrative Sciences and Arts, Arizona State University, Mesa, Arizona 85212, USA}
\affiliation{Department of Physics, Arizona State University, Tempe, Arizona 85287, USA}

\begin{abstract}
Using the Landau-level representation for the imaginary part of the photon polarization tensor, we derive an explicit expression for the dilepton production rate from a hot quark-gluon plasma in a quantizing background magnetic field. We study in detail the dependence of the production rate on the dilepton invariant mass and the transverse momentum at midrapidity. We also investigate in detail the angular dependence and ellipticity of dilepton emission. By comparing the result with the zero-field Born approximation, we find that the magnetic field leads to a strong enhancement of the dilepton rate at small values of the invariant mass ($M\lesssim\sqrt{|eB|}$). In the same kinematic region, the dilepton production is characterized by a sizable ellipticity. At large values of the dilepton invariant mass ($M\gtrsim\sqrt{|eB|}$), the role of the magnetic field decreases and the result approaches the isotropic zero-field Born rate. By investigating the dependence of ellipticity on the transverse momentum, we argue that the future measurements of dilepton rate in the region of small invariant masses can constrain the magnetic field produced in heavy-ion collisions.
\end{abstract}
%\pacs{12.38.Mh,25.75.-q,11.10.Wx,13.88+e}
\maketitle

\section{Introduction}

One of the main achievements of the heavy-ion research program at the Relativistic Heavy Ion Collider (RHIC) at Brookhaven and the Large Hadron Collider (LHC) at CERN is the discovery of hot quark-gluon plasma (QGP). The study of the fundamental properties of the corresponding deconfined state of matter is the broad goal of the ongoing program. The electromagnetic probes (i.e., photons and leptons) play a special role in this endeavor. Unlike the strongly interacting hadrons, they have long mean-free paths that greatly exceed the size of the fireballs created by the collisions. Thus, they carry invaluable information about the plasma directly to the detectors. Here, we will concentrate specifically on the dilepton emission. Recently, several measurements of the dilepton production were reported by both RHIC and LHC~\cite{Adamczyk:2015lme,Seck:2021mti,Acharya:2020gjz,Adam:2018tdm,Adam:2018qev,Adare:2015ila,Adamczyk:2015mmx,Adare:2009qk,Acharya:2018nxm}. 

The noncentral heavy-ion collisions produce the QGP  together with superstrong magnetic fields. For a fixed impact parameter, one can estimate the strength of the magnetic field during the early stages of the collision \cite{Skokov:2009qp,Deng:2012pc,Tuchin:2015oka,Guo:2019mgh}. Yet, it is uncertain how strong the field is during the later stages when the hot plasma forms and expands. If it remains relatively strong, the magnetic field can trigger anomalous phenomena, modify the flow of plasma, cause new types of collective modes, and affect the emission of particles. In this study, we explore how the magnetic field affects the dilepton production. Without the magnetic field, the thermal radiation from the QGP, the Drell--Yan process, and semileptonic decays of heavy quarks provide the dominant contributions to the dilepton rate in the intermediate range of the dilepton invariant masses. Notably, one often views the thermal part of the dilepton rate as a perfect thermometer of the QGP because the blue shift of the expanding medium does not modify it~\cite{Rapp:2014hha}. However, as we argue in this study, the dilepton emission is also a perfect magnetometer for the hot QGP with a strong background magnetic field. The signature effects of the magnetic field are the rate enhancement and strong anisotropy, whose measurements could provide valuable bounds on the field strength in the plasma produced by heavy-ion collisions.

The dilepton production rate for a hot QGP in the presence of a background magnetic field was studied previously by many authors using different approximations~\cite{Tuchin:2013bda,Sadooghi:2016jyf,Bandyopadhyay:2016fyd,Bandyopadhyay:2017raf,Ghosh:2018xhh,Islam:2018sog,Das:2019nzv,Ghosh:2020xwp,Chaudhuri:2021skc,Das:2021fma}. In particular, the dilepton rate was obtained in the equivalent photon flux approximation in Ref.~\cite{Tuchin:2013bda}, the lowest Landau level approximation in Refs.~\cite{Bandyopadhyay:2016fyd,Islam:2018sog}, and the weak field limit in Refs.~\cite{Bandyopadhyay:2017raf,Das:2019nzv}. Some results were obtained by using the Ritus method and the real-time formalism in Refs.~\cite{Sadooghi:2016jyf} and~\cite{Ghosh:2018xhh}, respectively. The corrections to the rate from the anomalous magnetic moment and the chiral condensate of quarks were studied in Refs.~\cite{Ghosh:2020xwp,Chaudhuri:2021skc} by using variants of the Nambu--Jona-Lasinio models. However, the kinematics was usually limited to the case of the vanishing perpendicular component of the dilepton momentum. Recently a more general result was presented in Ref.~\cite{Das:2021fma}. While the latter overlaps most closely with the current work, the angular dependence of the rate was not studied there. In this study, we extend the analysis by also calculating the ellipticity of the dilepton emission. 

The main goal of this paper is to deepen the theoretical understanding of dilepton production from a strongly magnetized hot QGP. We will investigate the dependence of the dilepton rate on the invariant mass $M$, the transverse momentum $k_{T}$ and, most importantly, the azimuthal angle $\phi$. The corresponding results will reveal a nontrivial ellipticity of dilepton production caused by the background magnetic field for a range of model parameters. As we will argue, such ellipticity carries information about the magnitude of the magnetic field at the early stages of heavy-ion collisions.

The starting point in our analysis will be the explicit expression for the imaginary part of the polarization tensor obtained in Ref.~\cite{Wang:2020dsr,Wang:2021ebh}. It leads to a relatively simple form for the differential dilepton production rate as a sum over the quark Landau levels. As we explain, one must treat the lepton states as plane waves. Accordingly, no sum over the lepton Landau levels appears in the rate. Such an approach is justified because (i) the mean free path of leptons is much longer than the QGP fireball, and (ii) the leptons themselves are detected far away from the magnetized QGP, where they are described naturally by plane waves. In simple terms, while the production occurs in a strongly magnetized plasma, the leptons have little chance of interacting with the plasma and get projected onto plane waves right after leaving the QGP. It is in a drastic contrast to several existing studies in the literature~\cite{Sadooghi:2016jyf,Bandyopadhyay:2016fyd,Bandyopadhyay:2017raf,Ghosh:2018xhh,Islam:2018sog,Das:2019nzv,Ghosh:2020xwp,Chaudhuri:2021skc}, where one uses the Landau levels for describing the final states of leptons. The corresponding treatment is misguided, however.

The paper is organized as follows. In Sec.~\ref{sec-2}, we start from the definition of dilepton rate in a magnetic field by using the one-loop photon polarization tensor in the Landau level representation. The corresponding numerical results are presented in Sec.~\ref{sec-3}. The summary of the main results and conclusions are given in Sec.~\ref{sec-4}. The Appendix at the end of the paper contains the expression for the magneto-optical conductivity.

\section{Formalism}
\label{sec-2}

\begin{figure}[t]
\centering
\includegraphics[width=0.33\textwidth]{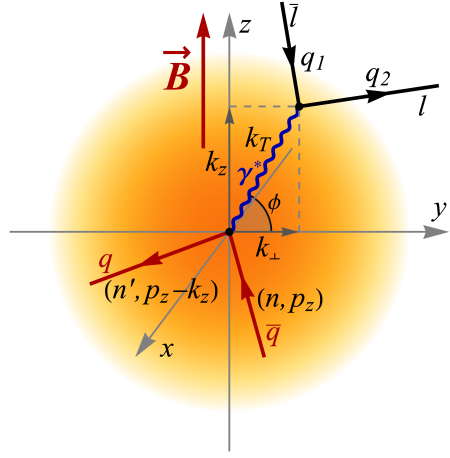}
\caption{A schematic illustration of the dilepton emission from a magnetized plasma at midrapidity. The transverse momentum of a virtual photon $\mathbf{k}_T$ lies in the $y$-$z$ plane. The azimuthal angle $\phi$ is measured from the reaction plane, i.e., the $x$-$y$ plane. The lepton momenta are $\mathbf{q}_1$ and $\mathbf{q}_2$.}
\label{fig:setup}
\end{figure}

The thermal emission of dileptons from a hot QGP is represented schematically by the Feynman diagram in Fig.~\ref{fig:setup}. We will focus on the local expanding plasma at midrapidity, assuming it is approximately thermalized. Without loss of generality, we will also assume that the magnetic field points in the $z$ direction and the beam is along the $x$ axis. In such a setup, the $x$-$y$ plane is the reaction plane.

Because of a quantizing background magnetic field, the quark and antiquark states in the QGP are characterized by the Landau-level quantum numbers, i.e., the integer indices $n$ and $n^\prime$ and the longitudinal momenta $p_z$ and $p_z^\prime \equiv p_z-k_z$, where $k_z$ is the longitudinal component of the virtual photon momentum. For simplicity, we will assume that the masses of both light quarks are the same, namely, $m_u = m_d = m = 5~\mbox{MeV}$. The flavor-dependent quark charges are $e_f = q_f e$, where $q_u = 2/3$, $q_d = -1/3$, and $e$ is the absolute value of the electron charge. 

Since the dilepton (virtual photon) is a neutral state, it is characterized by a well-defined four-momentum $K=(\Omega,\mathbf{k})$ with the magnitude of the spatial component given by $k = |\mathbf{k}|$. Given the rotational symmetry about the $z$ axis (coinciding with the magnetic field direction), it is sufficient to restrict the dilepton momentum to its transverse component $\mathbf{k}_T$ lying in the $y$-$z$ plane, which is perpendicular to the beam direction. We specify the direction of $\mathbf{k}_T$ in the transverse plane by the azimuthal angle $\phi$, which, by convention, is measured from the reaction plane, see Fig.~\ref{fig:setup}. In other words, the transverse components are $k_y=k_T \cos\phi$ and $k_z=k_T \sin\phi$. 

The dilepton rate (or, equivalently, the differential lepton multiplicity per unit spacetime volume) is given by~\cite{Weldon:1990iw}
\begin{equation}
d R_{l\bar{l}} =2\pi e^2 e^{-\beta \Omega} 
L_{\mu\nu}(Q_1,Q_2)\rho^{\mu\nu}(\Omega, \mathbf{k}) 
\frac{d^3 \mathbf{q}_1}{(2\pi)^3 E_1}\frac{d^3 \mathbf{q}_2}{(2\pi)^3 E_2} ,
\label{di-lepton}
\end{equation}
where the three-momenta and energies of the two leptons are denoted by $\mathbf{q}_i$ and $E_i$ with $i = 1, 2$, respectively. To the leading linear order in the electromagnetic coupling constant, the electromagnetic spectral function $\rho^{\mu\nu}(\Omega, \mathbf{k})$ is expressed in terms of the imaginary part of the photon polarization tensor as follows:
\begin{equation}
\rho^{\mu \nu}\left(\Omega, \mathbf{k}\right) = 
-\frac{1}{\pi} \frac{e^{\beta \Omega}}{e^{\beta \Omega}-1}  
\frac{\mbox{Im}\left[\Pi^{\mu \nu}\left(\Omega, \mathbf{k}\right)\right]}{K^{4}}.
\end{equation}
This spectral function includes the sum over all quark flavors, which is implicit in the photon polarization tensor. Moreover, the weights of the individual flavor contributions appear precisely as they should be in the dilepton production formula (\ref{di-lepton}). Its validity is reconfirmed indirectly by noting that the production of virtual photons is an intermediate process in the emission of dilepton pairs. We emphasize this point here because there are studies in the literature, see, for example Ref.~\cite{Bandyopadhyay:2016fyd}, which extract a factor of $q_f^2$ from the polarization tensor and introduce an effective coupling constant instead. Such an approach is unjustified since it changes the flavor weights and produces an incorrect expression for the rate.

In Eq.~(\ref{di-lepton}), the leptonic tensor for the final plane-wave states has the following explicit form:
\begin{equation}
L_{\mu \nu}(Q_1,Q_2)=\frac{1}{4} \sum_{\mathrm{spins}} \mbox{tr}\left[\bar{u}\left(Q_{2}\right) \gamma_{\mu} v\left(Q_{1}\right) \bar{v}\left(Q_{1}\right) \gamma_{\nu} u\left(Q_{2}\right)\right]=Q_{1 \mu} Q_{2 \nu}+Q_{1 \nu} Q_{2 \mu}-\left(Q_{1} \cdot Q_{2}+m_{l}^{2}\right) g_{\mu \nu},
\label{lepton-tensor}
\end{equation}
where $Q_2=K-Q_1$ due to the energy momentum conservation. We emphasize once again that the leptons are characterized by the usual four-momenta $Q_1$ and $Q_2$, which are the appropriate quantum numbers for the final states observed in a detector located far from the magnetized QGP. This may seem counterintuitive at first glance. Naively, since the leptons are produced inside the magnetized QGP, one might use the Landau-level eigenstates. As we explain below, the situation is more subtle.

While each lepton may be produced as a Landau-level state $|n_{\ell}\rangle$, it turns into a plane wave $|Q\rangle$ right after leaving the magnetized QGP. Quantum mechanically, the corresponding projection amplitude is given by $\langle n_{\ell} |Q\rangle$. Therefore, in order to calculate the production amplitude for the leptons with momenta $Q_1$ and $Q_2$ in the final state, one must sum over all intermediate (Landau-level) states. So, by making use of the completeness of such states, the corresponding amplitude can be written as follows:
\begin{equation}
\langle q_{n,p_z} , q_{n^\prime,p_z^\prime}| {\cal M} |Q_1 Q_2\rangle  = \sum_{n_{\ell_1}, n_{\ell_2}} \langle q_{n,p_z}\bar{q}_{n^\prime,p_z^\prime}| {\cal M} |n_{\ell_1}, n_{\ell_2}\rangle \langle n_{\ell_1} |Q_1\rangle
\langle n_{\ell_2} |Q_2\rangle,
\label{momentum-states}
\end{equation}
where $q_{n,p_z}$ and $q_{n^\prime,p_z^\prime}$ represent the initial quark states in the magnetized plasma. In essence, while the amplitude of the dilepton production inside the magnetized QGP is $\langle q_{n,p_z}\bar{q}_{n^\prime,p_z^\prime}| {\cal M} |n_{\ell_1}, n_{\ell_2}\rangle$, the amplitude associated with the detector measurement is given by  $\langle q_{n,p_z} , q_{n^\prime,p_z^\prime}| {\cal M} |Q_1 Q_2\rangle$.

The relation in Eq.~(\ref{momentum-states}) formally justifies the use of plane waves in the expression for the dilepton rate (\ref{di-lepton}). One should note, however, that the use of quantum-state projections alone would not be sufficient if the leptons had a high probability of scattering inside the plasma. Fortunately, the corresponding probability is very small since the scattering mean free path of leptons is much larger than the plasma regions produced by heavy-ion collisions. Therefore, Eq.~(\ref{momentum-states}) is indeed a good approximation.

By using the explicit form of the leptonic tensor (\ref{lepton-tensor}), we rewrite the differential dilepton production rate as follows:
\begin{equation}
\label{rate_0}
d R_{l\bar{l}} =- \frac{\alpha}{8\pi^5} 
 n_{B}\left(\Omega\right)  L_{\mu\nu}(Q_1,Q_2)
 \frac{\mbox{Im}\left[\Pi^{\mu \nu}\left(\Omega, \mathbf{k}\right)\right]}{K^{4}}
\frac{d^3 \mathbf{q}_1}{E_1}\frac{d^3 \mathbf{q}_2}{E_2} ,
\end{equation}
where $\alpha \equiv e^2/(4\pi)= 1/137$ is the fine structure constant and $n_B(\Omega) = (e^{\Omega/T}-1)^{-1}$ is the Bose-Einstein distribution function. In terms of the total four-momentum $K$ of the lepton pair, the corresponding differential rate reads \cite{Weldon:1990iw}
\begin{equation}
\label{rate_1}
\frac{d R_{l\bar{l}}}{d^{4} K}=\frac{ \alpha}{12 \pi^4} \frac{n_{B}\left(\Omega\right)}{K^{2}}\left(1+\frac{2 m_{l}^{2}}{K^{2}}\right)\left(1-\frac{4 m_{l}^{2}}{K^{2}}\right)^{1 / 2}  \mbox{Im}\left[\Pi_{\mu}^{\mu}\left(\Omega, \mathbf{k}\right)\right] .
\end{equation}
For the purposes of this study, the lepton masses $m_{l}$ can be set to zero. This is justified since the dilepton energies and momenta of interest will be much larger than the lepton masses. Then, the expression for the rate further simplifies, i.e.,  
\begin{equation}
\label{rate_general}
\frac{d R_{l\bar{l}}}{d^{4} K}=\frac{ \alpha}{12 \pi^4} \frac{n_{B}\left(\Omega\right)}{M^{2}} \mbox{Im}\left[\Pi_{\mu}^{\mu}\left(\Omega, \mathbf{k}\right)\right],
\end{equation}
where $M^2 =K^2 \equiv \Omega^2 - k_\perp^2 -k_z^2$ is the square of the invariant mass of the lepton pair. By definition, $k_\perp= \sqrt{k_x^2+k_y^2}$ is the magnitude of the momentum component perpendicular to the magnetic field. Note that $\Omega =\sqrt{M^2+k_\perp^2 +k_z^2}$ and $d^{4} K = MdM k_T dk_T dy d\phi$, where $k_T =\sqrt{k_y^2+k_z^2}$ is the transverse momentum (with respect to the beam direction) and $y = \frac{1}{2}\ln\frac{\Omega+k_x}{\Omega-k_x}$ is the rapidity.

To obtain the explicit expression for the dilepton rate in a strongly magnetized QGP, we use the imaginary part of the Lorentz-contracted photon polarization tensor  that was derived recently in Refs.~\cite{Wang:2020dsr,Wang:2021ebh}. The final result reads
\begin{eqnarray}
\frac{d R_{l\bar{l}}}{d^{4} K}&=&\frac{\alpha^2 N_c}{48 \pi^5} \frac{n_{B}\left(\Omega\right)}{M^{2}} 
\sum_{f=u, d} \frac{q_f^2}{\ell_f^4 } \Bigg[ \sum_{n=0}^{\infty} 
\frac{g_0(n)\theta\left(\sqrt{M^2+k_\perp^2}-k_{+}^{f}\right)}{\sqrt{(M^2+k_\perp^2) \left[M^2+k_\perp^2-(k_{+}^{f})^2\right]} }\mathcal{F}_{n, n}^f (\xi)
\nonumber \\
&-&2\sum_{n>n^\prime}^{\infty}  
\frac{g(n, n^{\prime}) 
\left[
\theta\left(k_{-}^{f}-\sqrt{M^2+k_\perp^2}\right)
-\theta\left(\sqrt{M^2+k_\perp^2}-k_{+}^{f}\right) \right]
 }{\sqrt{ \left[( k_{-}^{f} )^2 -(M^2+k_\perp^2) \right]\left[ (k_{+}^{f})^2-(M^2+k_\perp^2)\right] } } 
\mathcal{F}_{n, n^{\prime}}^f(\xi)   \Bigg], 
\label{rate_f}
\end{eqnarray}
where $\ell_f = \sqrt{1/|e_f B|}$ is the flavor-specific magnetic length. The Heaviside step-function is denoted by $\theta(x)$ and the Landau-level threshold momenta are given by $k_{\pm}^f = \left|\sqrt{m^2+2n|e_f B|}\pm\sqrt{m^2+2n^{\prime}|e_f B|}\right|$. The explicit expressions for functions $g(n, n^{\prime}) $ and $g_0(n)$ are given by 
\begin{eqnarray}
g(n, n^{\prime}) &=& 2-\sum_{s_1,s_2=\pm}
n_F\left(\frac{\Omega}{2}  +s_1 \frac{\Omega(n-n^{\prime})|e_fB|}{M^2+k_\perp^2}+\frac{s_2  |k_z|}{2(M^2+k_\perp^2)}\sqrt{ \left(M^2+k_\perp^2-(k_{-}^{f})^2 \right)\left(M^2+k_\perp^2- (k_{+}^{f})^2\right)} 
\right), \\
g_0(n) &=& g(n, n)  =2 - 2\sum_{s=\pm}
n_F\left(\frac{\Omega}{2} +\frac{s  |k_z|}{2\sqrt{M^2+k_\perp^2}}\sqrt{M^2+k_\perp^2 - 4\left(m^2+2n|e_f B|\right)}
\right),
\end{eqnarray}
where $n_F(\Omega) = (e^{\Omega/T}+1)^{-1}$ is the Fermi-Dirac distribution function.

Note that the dependence of the rate on the perpendicular component of the virtual photon momentum $k_\perp$ enters in Eq.~(\ref{rate_f}) not only explicitly but also via $\mathcal{F}^f_{n, n^{\prime}} (\xi)$, which is a function of $\xi =(k_{\perp}\ell_f)^{2}/2$. The explicit form of the corresponding function reads
\begin{equation}
\mathcal{F}^f_{n, n^{\prime}} (\xi)
= 8\pi \left(n+n^{\prime}+m^2\ell_f^2\right)\left[\mathcal{I}_{0,f}^{n,n^{\prime}}(\xi)+\mathcal{I}_{0,f}^{n-1,n^{\prime}-1}(\xi) \right]
+8\pi  \left(\frac{M^2\ell_f^2}{2}   -(n+n^{\prime})\right)
\left[\mathcal{I}_{0,f}^{n,n^{\prime}-1}(\xi)+\mathcal{I}_{0,f}^{n-1,n^{\prime}}(\xi) \right],
\end{equation}
where functions $\mathcal{I}_{0,f}^{n,n^{\prime}}(\xi)$ are defined in terms of the generalized Laguerre polynomials \cite{Gradshtein} as follows:
\begin{equation}
\label{Ixl}
\mathcal{I}_{0,f}^{n,n'}(\xi) = (-1)^{n+n^\prime} \ell_f^2  e^{-\xi} L_{n}^{n^\prime-n}\left(\xi\right) 
L_{n^\prime}^{n-n^\prime}\left(\xi\right).
\end{equation}
In Sec.~\ref{sec-3} below, we will use the general expression in Eq.~(\ref{rate_f}) to calculate the dilepton rate numerically and study its dependence on the kinematic parameters. 

When studying the dilepton rate in QGP with a strong background magnetic field, it is useful to compare the results with the corresponding rate in the zero-field limit. In the Born approximation, such a rate is given by~\cite{Cleymans:1986na}
\begin{equation}
\label{rate_B0_Born}
\frac{dR_{l\bar{l},{\rm Born}}}{d^4 K} = \frac{5 \alpha^2 T}{18\pi^4 |\bm{k}|}n_B(\Omega)\ln\left(\frac{\cosh\frac{\Omega+|\bm{k}|}{4T}}{\cosh\frac{\Omega-|\bm{k}|}{4T}}\right)  ,
\end{equation}
where the massless quarks and leptons are assumed.

It is interesting to note that the result for the dilepton rate simplifies considerably in the zero-momentum limit. As one can see, the corresponding result is proportional to the trace of the magneto-optical conductivity tensor $\sigma_{ij} (\Omega)$ evaluated at $\Omega=M$, i.e.,
\begin{equation}
\label{rate_conductivity}
\left.\frac{d R_{l\bar{l}}}{d^{4} K}\right|_{|\bm{k}|\to 0}\simeq\frac{ \alpha}{12 \pi^4} \frac{n_{B}\left(M\right)}{M}  \left[\sigma_\parallel(M)+ 2\sigma_\perp(M)\right],
\end{equation}
where $\sigma_\parallel(\Omega)$ and $\sigma_\perp(\Omega)$ are the longitudinal and transverse (with respect to the magnetic field) components of magneto-optical conductivity, respectively. For a generic strongly magnetized relativistic plasma, both components of conductivity were calculated in Ref.~\cite{Wang:2021ebh}. To extend the corresponding expressions to the case of QGP, one must include the contributions of all quark flavors, see Appendix~\ref{AppA}. 

As we show in Appendix~\ref{AppA}, both longitudinal and transverse components of optical conductivity reduce to the same expression in the limit of the vanishing magnetic field:
\begin{equation}
\left. \sigma_\parallel(\Omega)\right|_{B\to 0} = \left.\sigma_\perp(\Omega)\right|_{B\to 0}   \simeq  \frac{ \alpha N_{c} (q_u^2+q_d^2)}{3} \Omega \,\tanh\left(\frac{\Omega}{4T}\right) ,
\end{equation} 
where we assumed the vanishing quark masses. By substituting the optical conductivity into Eq.~(\ref{rate_conductivity}), we obtain
\begin{equation}
\label{rate_conductivity_B0}
\left.\frac{d R_{l\bar{l}}}{d^{4} K}\right|_{|\bm{k}|\to 0, B\to 0}\simeq\frac{5 \alpha^2}{36 \pi^4}  n_{B}\left(M\right) \tanh\left(\frac{M}{4T}\right).
\end{equation}
As expected, this result agrees with the Born rate  (\ref{rate_B0_Born}) in the limit $|\bm{k}| \to 0$.

\section{Numerical Results}
\label{sec-3}

In this section, we study numerically the dependence of the dilepton production rate in Eq.~(\ref{rate_f}) on the main kinematic parameters, namely the invariant mass $M$, the transverse momentum $k_T$, and the azimuthal angular coordinate $\phi$. For simplicity, we will limit consideration only to the case of midrapidity by setting $k_x = 0$. We will also use the angular dependence to extract the ellipticity of the dilepton emission.

To probe different regimes of strongly magnetized QGP, we will calculate the rate for two representative values of the magnetic field strength, i.e., $|eB|=m_\pi^2$ and $|eB|=5m_\pi^2$, where $m_{\pi}\approx 0.135~\mbox{GeV}$ is the (neutral) pion mass. In conventional units, the corresponding values of the field are $B\approx 3.08\times 10^{18}~\mbox{G}$ and $B\approx 1.54\times 10^{19}~\mbox{G}$, respectively. To understand how the magnetic field interplays with the thermal effects in the QGP, we will also consider two different representative temperatures, i.e., $T=0.2~\mbox{GeV}$ and $T=0.35~\mbox{GeV}$. Roughly speaking, one may view them as the temperatures at the early and late stages of the QGP produced in heavy-ion collisions.

\subsection{Dilepton rate dependence on $M$ and $k_T$}

The dilepton rate is anisotropic in the presence of a magnetic field. This is reconfirmed by the general expression in Eq.~(\ref{rate_f}), which has a nontrivial dependence on $k_\perp=k_T \cos\phi$ and, thus, on the angular coordinate $\phi$. Before addressing the subtleties of the angular dependence, it is instructive to define the total differential rate, i.e.,
\begin{equation}
\frac{dR_{l\bar{l}}}{MdM k_Tdk_T dy}=  \int_0^{2\pi} d\phi \frac{d R_{l\bar{l}} }{d^{4} K} .
\label{rate-all-phi}
\end{equation}
To calculate the differential rate in the whole range of the angular coordinate $\phi$ more efficiently, we use the spatial symmetries in the magnetized plasma at midrapidity. For the setup in Fig.~\ref{fig:setup}, there are two relevant symmetries. The first one is a subgroup of spatial rotations about the $z$ axis, which remains unbroken in the presence of a constant background magnetic field. The other is the mirror reflection in the $x$-$y$ plane, which is the consequence of the magnetic field being a parity-even axial vector. In application to the rate, we utilize the symmetry under the rotation by angle $\pi$ about the $z$ axis to obtain $d R_{l\bar{l}}/d^{4} K (\pi-\phi) =d R_{l\bar{l}}/d^{4} K (\phi)$. In turn, the mirror reflection symmetry leads to $d R_{l\bar{l}}/d^{4} K (-\phi) =d R_{l\bar{l}}/d^{4} K (\phi)$. Therefore, the knowledge of the differential rate in the first quadrant, i.e., $0\leq \phi < \pi/2$, is sufficient to obtain the results in the other three quadrants too.

To compile the numerical dependence of the differential rate on $\phi$, we use a large set ($n_{\phi}=1001$) of equidistant angular coordinates in the first quadrant, ranging from $\phi_{\rm min}=10^{-4}\frac{\pi}{2}$ to $\phi_{\rm max}=\frac{\pi}{2} - \phi_{\rm min}$, with a small discretization step $\Delta \phi \simeq 10^{-3}(\phi_{\rm max}-\phi_{\rm min})$. Note that there is nothing special about the rates at the limiting values of the angular coordinate $\phi=0$ and $\phi=\pi/2$, provided no accidental Landau-level thresholds appear at those values. Originally, we avoided the limiting values of $\phi$ out of an abundance of caution. However, the numerical values of the rates at $\phi=0$ and $\phi=\pi/2$ are nearly indistinguishable from those at $\phi_{\rm min}$ and $\phi_{\rm max}$, respectively. The total integrated rate in Eq.~(\ref{rate-all-phi}) is obtained approximately by calculating the sum of the individual contributions, each multiplied by the weight factor $\Delta \phi$. (To account for the contributions from all four quadrants, the summation result must be also multiplied by a factor of $4$.)

When calculating the differential rate in Eq.~(\ref{rate_f}) numerically, one of the challenging tasks is the sum over Landau levels. The  sum needs to be truncated at a finite but sufficiently large number of terms $n_{\rm max}$. The estimate for $n_{\rm max}$ is the same as that in the photon emission study but expressed in terms of $\Omega=\sqrt{M^2+k_T^2}$ instead of $k_T$  \cite{Wang:2021ebh}. Indeed, when $\Omega$ is small compared to the magnetic field scale $\sqrt{|eB|}$, large separations between low-lying Landau levels strongly suppresses contributions of quark states with small Landau indices. In this regime, we find that many Landau levels, with indices up to about $n_{\rm max}\simeq |eB|/\Omega^{2}$, contribute. In the opposite limit of large $\Omega$, one requires a large enough cutoff in quark energies merely to open the phase space for dilepton production. The corresponding requirement translates into the following estimate for the Landau index cutoff: $n_{\rm max}\simeq \Omega^{2}/|eB|$. For simplicity, we use a fixed value $n_{\rm max}=1000$ in the numerical calculations. It is sufficiently large to render reliable results for the rate in a wide range of energies from about $\Omega_{\rm{min}} \simeq 0.02~\mbox{GeV}$ to about $\Omega_{\rm{max}}\simeq 2~\mbox{GeV}$.

\begin{figure}[t]
\centering
\includegraphics[width=0.46\textwidth]{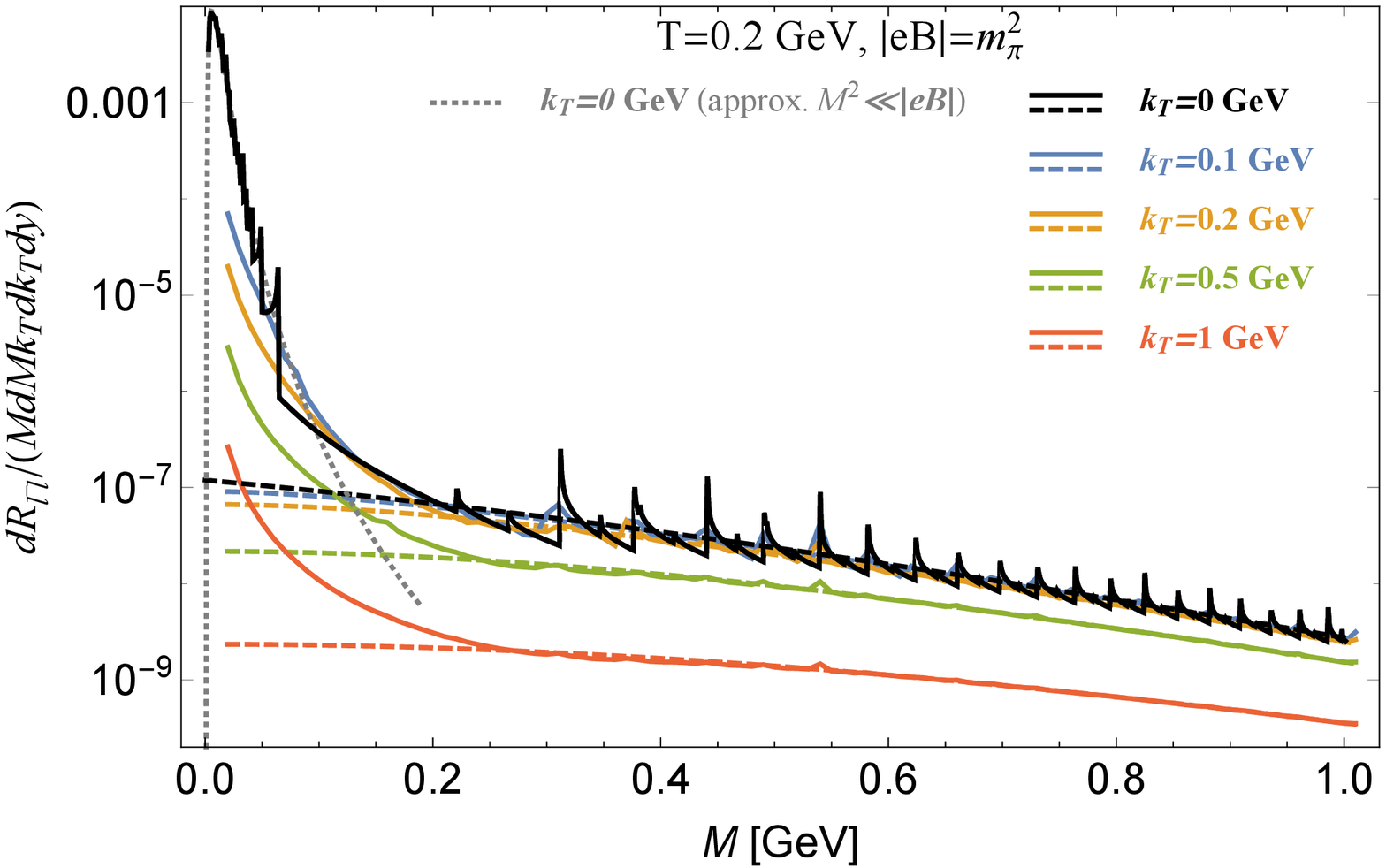}
\hspace{0.05\textwidth}
\includegraphics[width=0.46\textwidth]{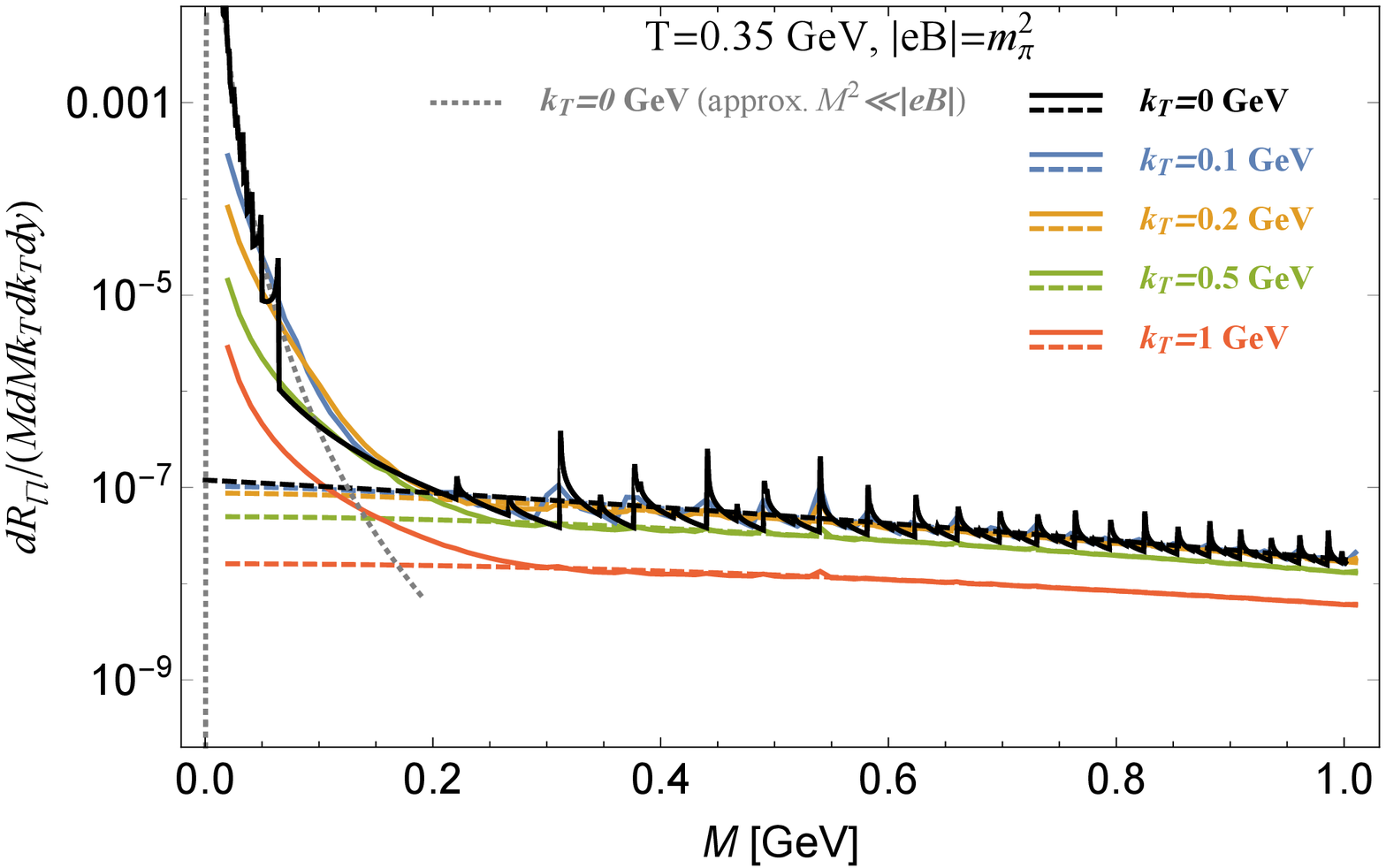}
\includegraphics[width=0.46\textwidth]{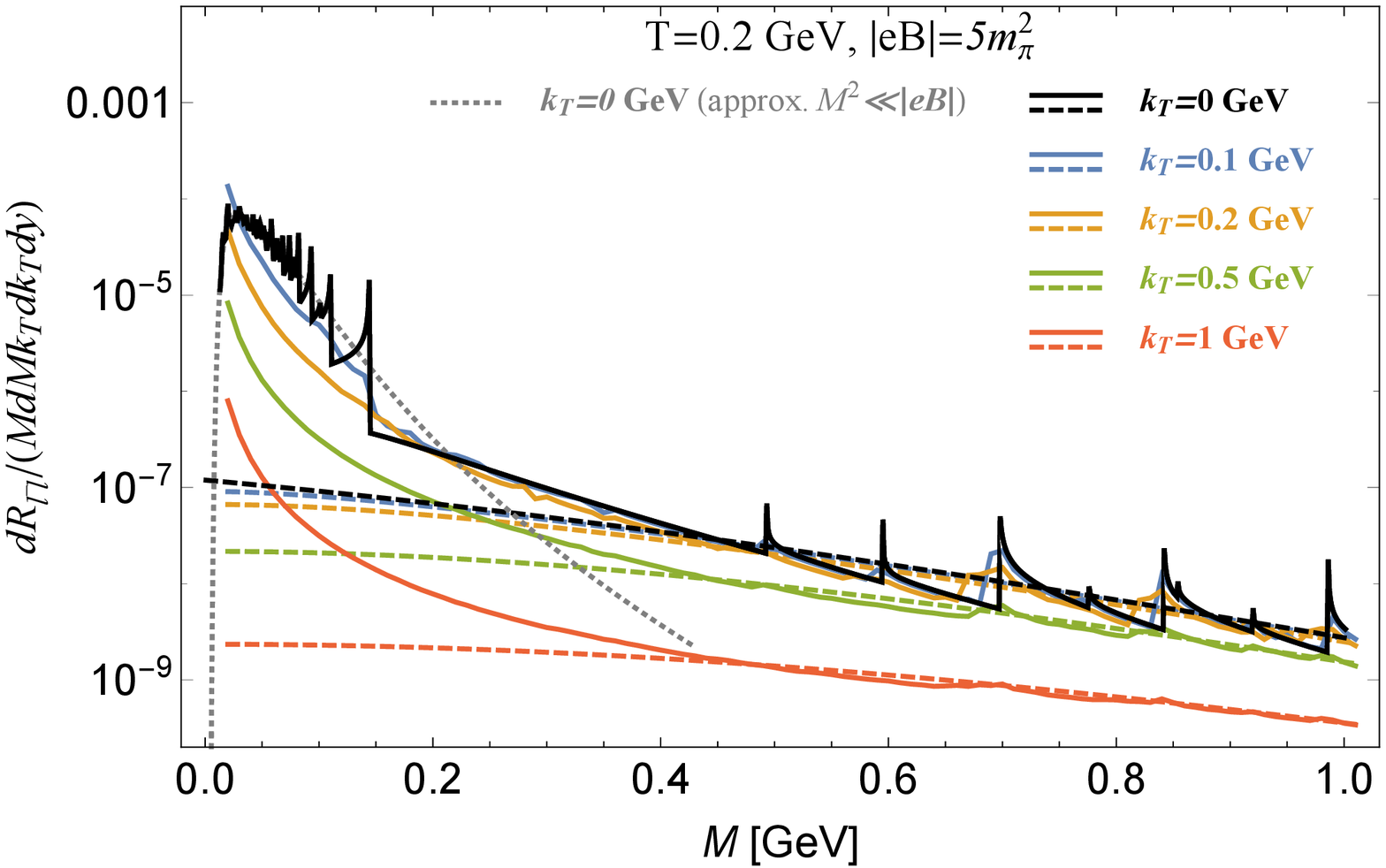}
\hspace{0.05\textwidth}
\includegraphics[width=0.46\textwidth]{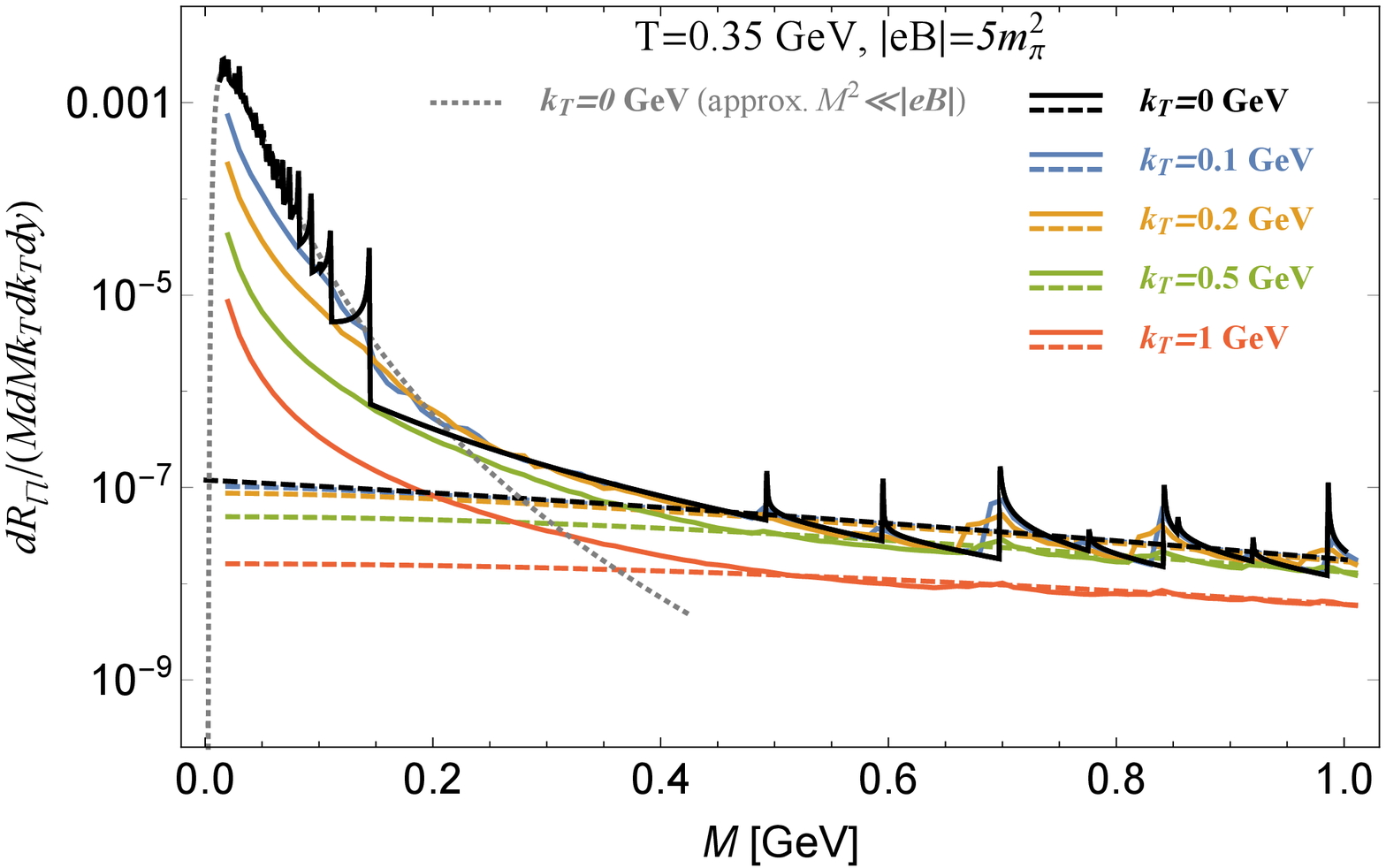}
\caption{The integrated dilepton rate as a function of the dilepton invariant mass $M$ for several fixed values of the transverse momentum $k_T = 0, 0.1, 0.2, 0.5, 1~\mbox{GeV}$, two values of the temperature, i.e., $T = 0.2~\mbox{GeV}$ (left panels) and $T = 0.35~\mbox{GeV}$ (right panels), and two values of the magnetic field, i.e., $|eB| = m_{\pi}^2$ (top panels) and $|eB| = 5m_{\pi}^2$ (bottom panels). For comparison, the dashed lines represent the zero-field Born rate and the gray dotted lines show the approximate rate at $k_T = 0$ and $M\ll \sqrt{|eB|}$, given in Eq.~(\ref{rate_conductivity_kT0-small-M}).}
\label{Fig:rate-vs-M}
\end{figure}

The dependence of the rate (\ref{rate-all-phi}) on the invariant mass is presented in Fig.~\ref{Fig:rate-vs-M}. The corresponding data was calculated for the whole range of invariant masses between $M_{\rm min}=0.02~\mbox{GeV}$ and $M_{\rm max}=1~\mbox{GeV}$, using the discretization step $\Delta M=0.01~\mbox{GeV}$. Individual panels show the results for two representative choices of temperature, i.e., $T = 0.2~\mbox{GeV}$ (two left panels) and $T = 0.35~\mbox{GeV}$ (two right panels), and two values of the magnetic field, i.e., $|eB| = m_{\pi}^2$ (two top panels) and $|eB| = 5m_{\pi}^2$ (two bottom panels). Each panel contains the rates for the same set of fixed values of the transverse momenta, i.e., $k_T = 0$ (black), $k_T = 0.1~\mbox{GeV}$ (blue), $k_T = 0.2~\mbox{GeV}$ (orange), $k_T = 0.5~\mbox{GeV}$ (green), and $k_T = 1~\mbox{GeV}$ (red). For comparison, we also show the zero-field Born rate (dashed lines), which is defined by Eq.~(\ref{rate_B0_Born}).

As expected, the Landau-level quantization of the quark spectrum produces a nonsmooth dependence of the rate on the invariant mass. The origin of many sharp spikes in the dilepton rate is the same as in the photon emission, discussed in detail in Ref.~\cite{Wang:2021ebh}. They show up whenever the center-of-mass energy crosses one of the numerous Landau-level thresholds, leading to a sudden change of the kinematic phase space for the dilepton production. As seen in Fig.~\ref{Fig:rate-vs-M},  the spiky behavior is pronounced most strongly at low transverse momenta. If such signature quantization effects could be observed in an experiment, they would provide the unambiguous proof of a strong background magnetic field in the plasma. The chances of that are small, however. For one thing, the Landau-level threshold effects should smooth out when strong quark interactions are taken into account \cite{Wang:2021ebh}. Also, additional smoothing will necessarily come from the time evolution of the background magnetic field in the QGP. 

By disregarding the spiky behavior, it is still instructive to compare the overall profiles of dilepton rates in a strongly magnetized plasma with the benchmark zero-field Born rate. We find that the rates at $B\neq 0$ remain about the same on average as those at $B=0$ when the invariant mass is sufficiently large, i.e., $M\gtrsim \sqrt{|eB|}$. On the other hand, the magnetic field has a dramatic effect on the dilepton production in the region of small invariant masses, i.e., $M\lesssim \sqrt{|eB|}$, where the rates are strongly enhanced. As seen in the four panels of Fig.~\ref{Fig:rate-vs-M}, the rate can increase by several orders of magnitude when $M$ decreases only by half. The same qualitative dependence on the invariant mass remains robust for a wide range of transverse momenta and other plasma parameters, including different temperatures and magnetic fields.
 
As we see from Fig.~\ref{Fig:rate-vs-M}, the dilepton rate is on-average a decreasing function of the invariant mass $M$. The rate for the vanishing (sufficiently small) $k_T$ is an exception. Its nonmonotonic dependence at small $M$ will be discussed below. Generically, the rate in the magnetized QGP approaches the zero-field Born result (\ref{rate_B0_Born}) when the invariant mass is sufficiently large (i.e.,  $M\gg \sqrt{|eB|}$). By comparing the plots in Fig.~\ref{Fig:rate-vs-M} for different values of the transverse momenta, we see that the dilepton rate tends to decrease with increasing $k_T$. As one can verify, the suppression of the rate at large $M$ or $k_T$ (or both) comes primarily from the overall Bose distribution function $n_B(\Omega)$ in Eq.~(\ref{rate_f}).
 
It is interesting to note that the dilepton rate at $k_T=0$ is nonmonotonic, with a maximum appearing at a small but nonzero invariant mass. The corresponding maxima are seen in Fig.~\ref{Fig:rate-vs-M} for three out of the four sets of model parameters. (The maximum is present but outside the plot range when $T=0.35~\mbox{GeV}$ and $|eB|= m_\pi^2$.) We can show that such a behavior is universal for the vanishing (or sufficiently small) $k_T$. Indeed, by using Eq.~(\ref{rate_conductivity}) and the optical conductivities, obtained in Appendix~\ref{AppA}, we derive the following analytical expression for the dilepton rate at $k_T=0$:
\begin{equation}
\label{rate_conductivity_kT0-small-M}
\left.\frac{d R_{l\bar{l}}}{d^{4} K}\right|_{|\bm{k}|= 0}\simeq \sum_{f=u,d}\frac{\alpha^2 N_{c}  q_f^2 |e_fB|^3  \exp\left(-\frac{M}{2T}\right) }{9 \pi^4 M^6 \left[\cosh\left(\frac{M}{2T}\right) +\cosh\left(\frac{|e_fB|}{TM}\right) \right]},
\end{equation}
which is valid when $M\ll \sqrt{|eB|}$. It describes the regime when a large separation between the low-lying Landau levels of quarks suppresses the production of dileptons with a small invariant mass (note that $\Omega = M$ at $k_T=0$). Instead, transitions between quark states with sufficiently large but closely lying energies dominate the rate. The corresponding phase space includes many transitions between Landau levels with indices up to $n_{\rm max}\lesssim |e_fB|/(2\Omega^2)$, see Appendix~\ref{AppA}. Note that the rate in Eq.~(\ref{rate_conductivity_kT0-small-M}) describes the strong-field limit where the lowest Landau level approximation is completely inapplicable.

The maximum value of the rate in Eq.~(\ref{rate_conductivity_kT0-small-M}) is achieved at $M_{\rm max}\simeq \sqrt{(6T)^2+2 |e_d B|}-6T$. Around the maximum, the dominant contribution comes from the down quarks. For the magnetic fields and temperatures considered in this study, the value of $ |e_d B|$ is at least an order of magnitude smaller than $(6T)^2$. Thus, one can use the expansion in powers of small $|e_d B|$ to obtain an approximate location of the peak: $M_{\rm max}\simeq |e_d B|/(6T)$.  Furthermore, we find that the corresponding rate is a factor of order of $(3T)^6/|e_d B|^3$ larger than the zero-field limit in Eq.~(\ref{rate_conductivity_B0}).

\subsection{Angular dependence of the dilepton rate}

\begin{figure}[t]
\centering
\includegraphics[width=0.46\textwidth]{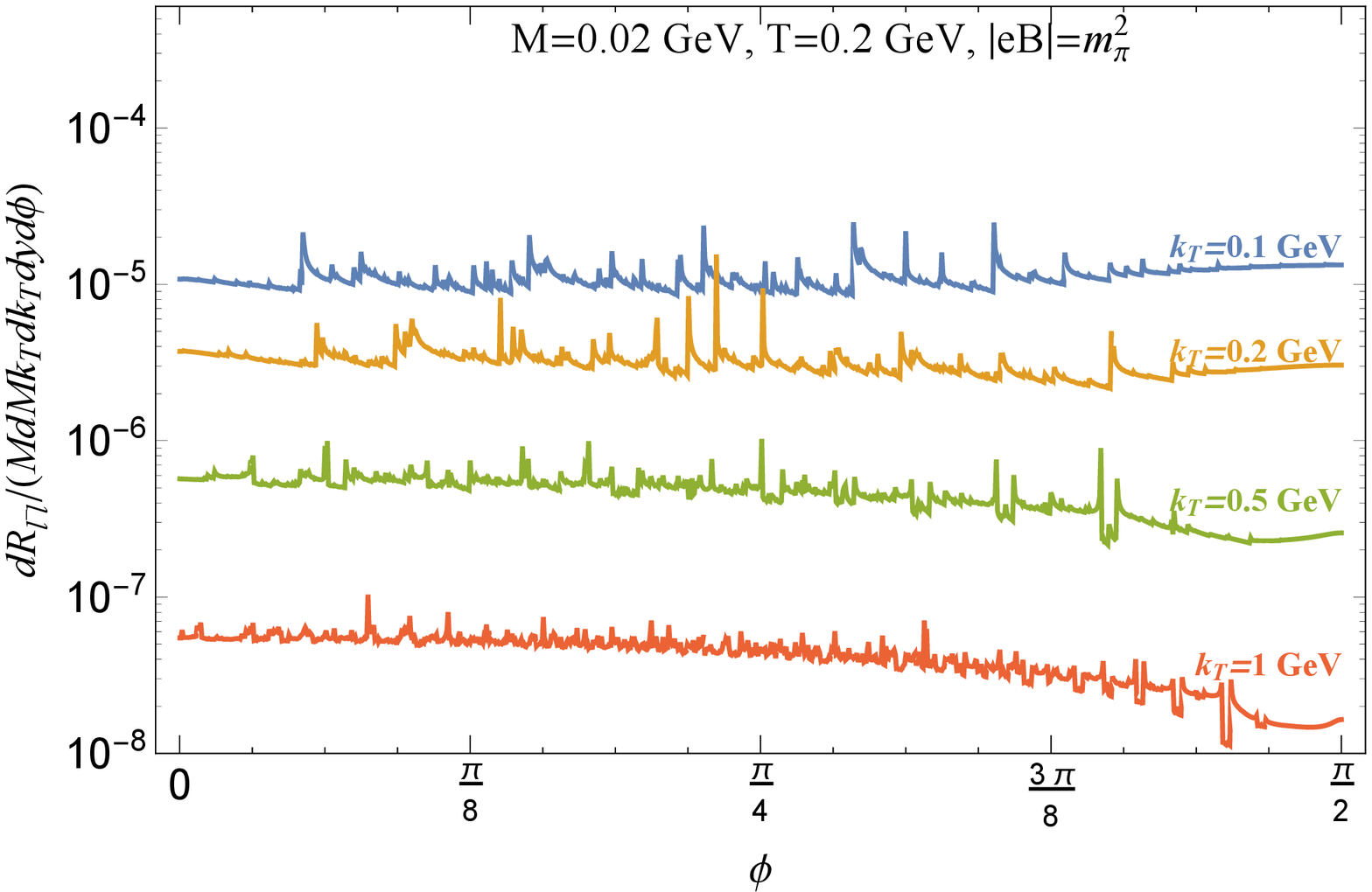}
\hspace{0.05\textwidth} 
\includegraphics[width=0.46\textwidth]{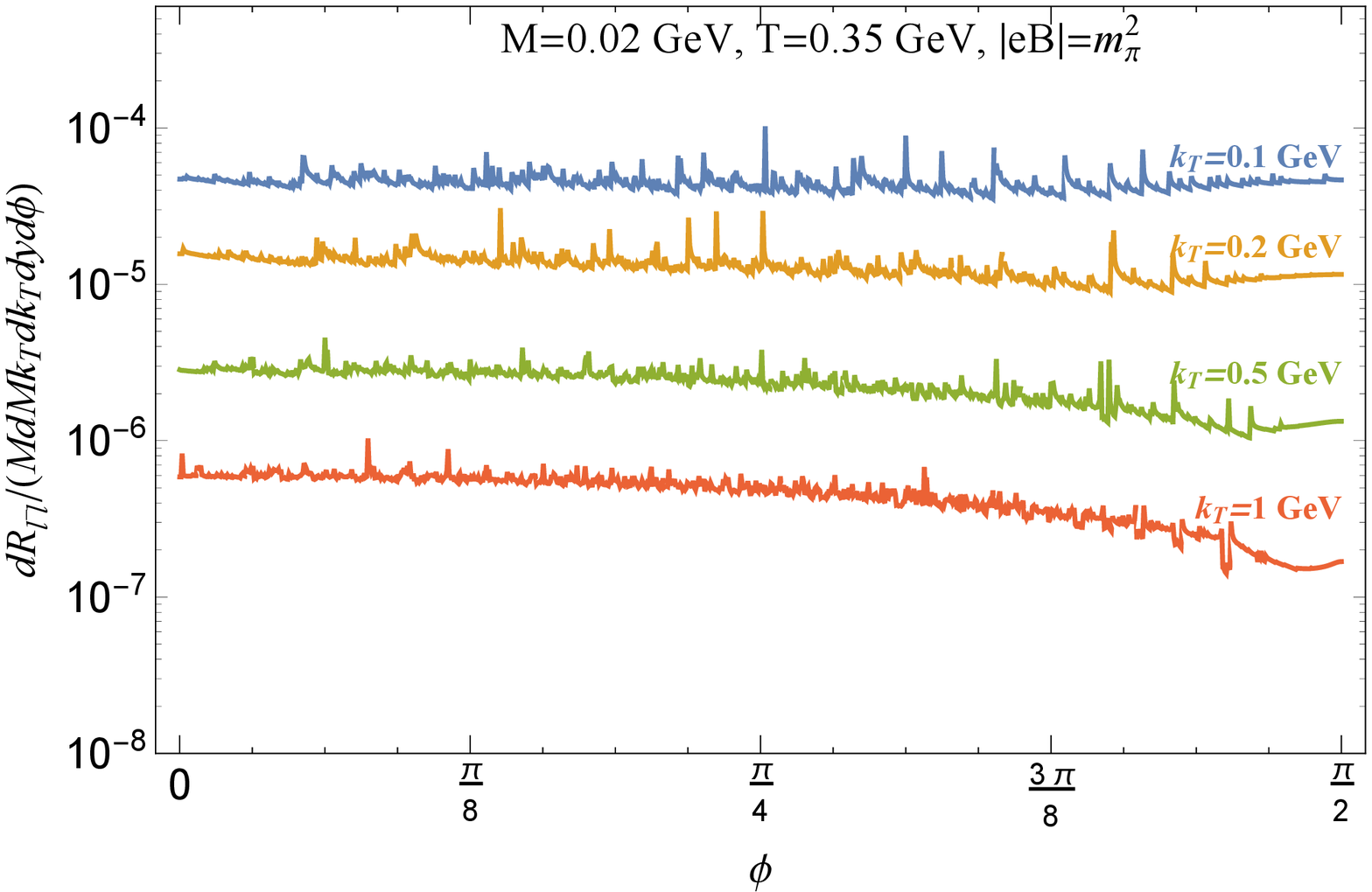}
\includegraphics[width=0.46\textwidth]{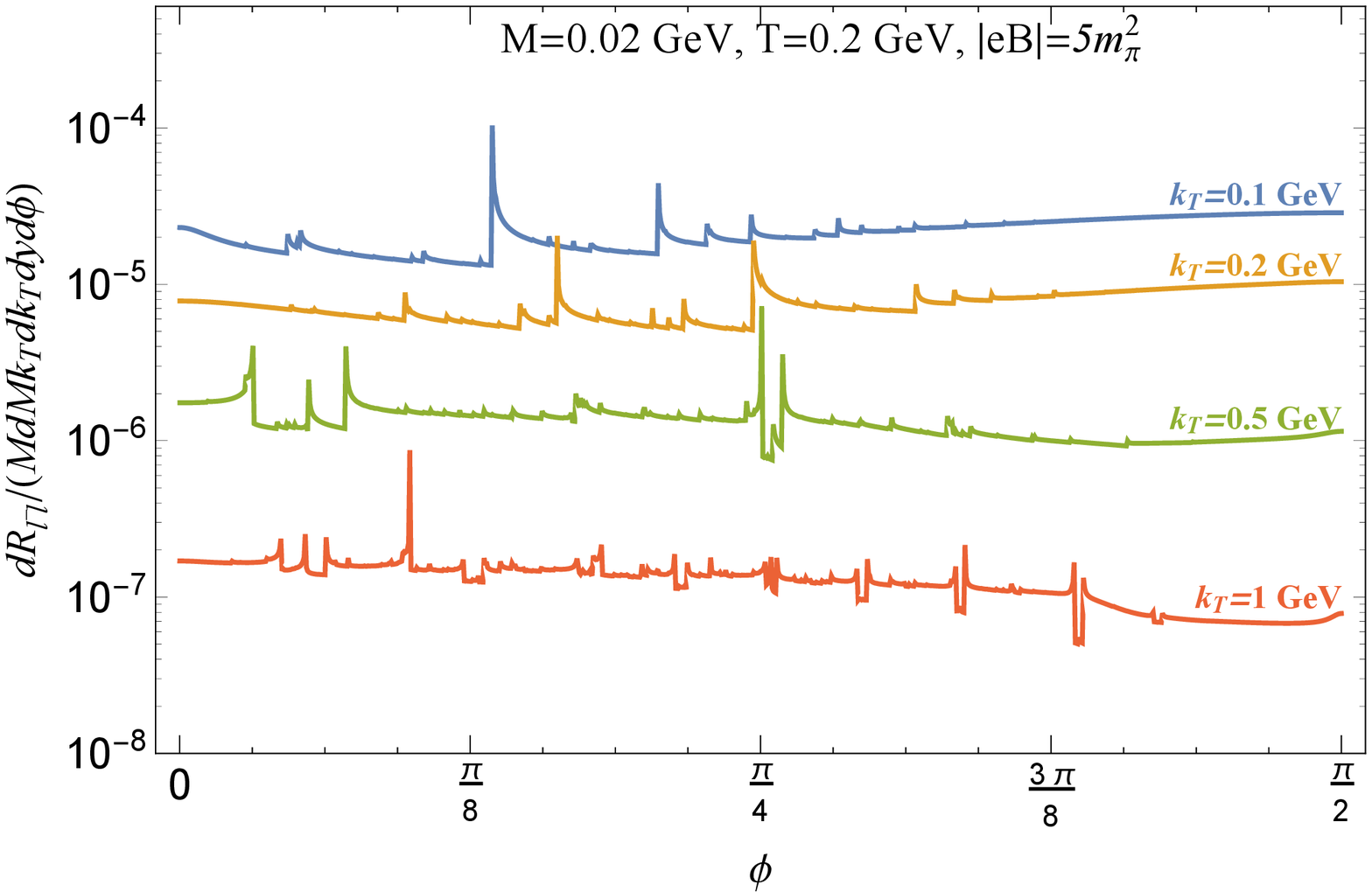}
\hspace{0.05\textwidth}
\includegraphics[width=0.46\textwidth]{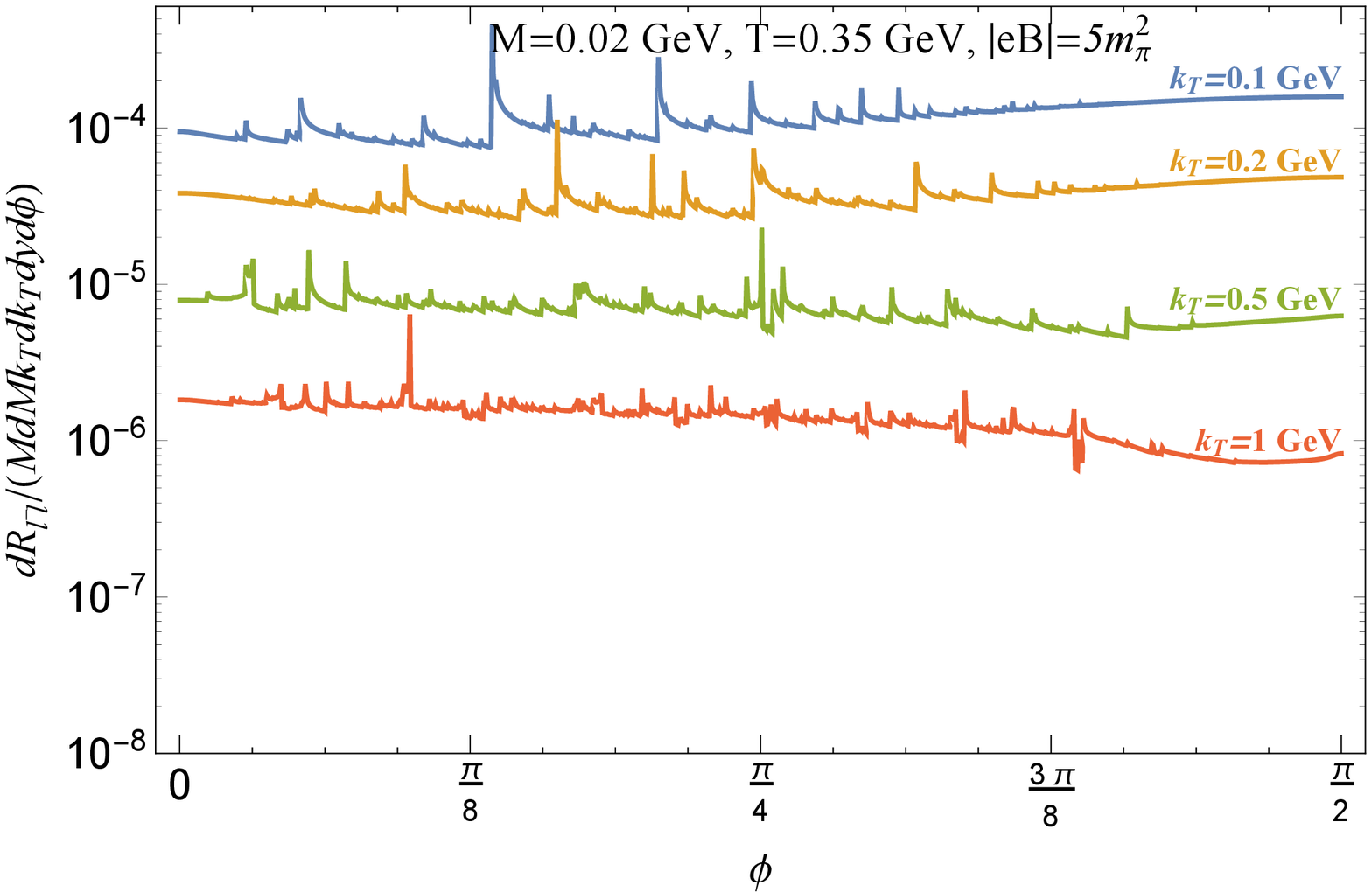}
\caption{The dilepton production rate as a function of the azimuthal angle $\phi$ for a fixed invariant mass $M = 0.02~\mbox{GeV}$, two temperatures, i.e., $T = 0.2~\mbox{GeV}$ (left panels) and $T = 0.35~\mbox{GeV}$ (right panels), and two magnetic fields, i.e., $|eB| = m_{\pi}^2$ (top panels) and $|eB| = 5m_{\pi}^2$ (bottom panels). Each panel shows the rates for four fixed transverse momenta, i.e., $k_T = 0.1~\mbox{GeV}$ (blue), $k_T = 0.2~\mbox{GeV}$ (orange), $k_T = 0.5~\mbox{GeV}$ (green), and $k_T = 1~\mbox{GeV}$ (red).}
\label{Fig:rate-vs-phi_M002}
\end{figure}

\begin{figure}[t]
\centering
\includegraphics[width=0.46\textwidth]{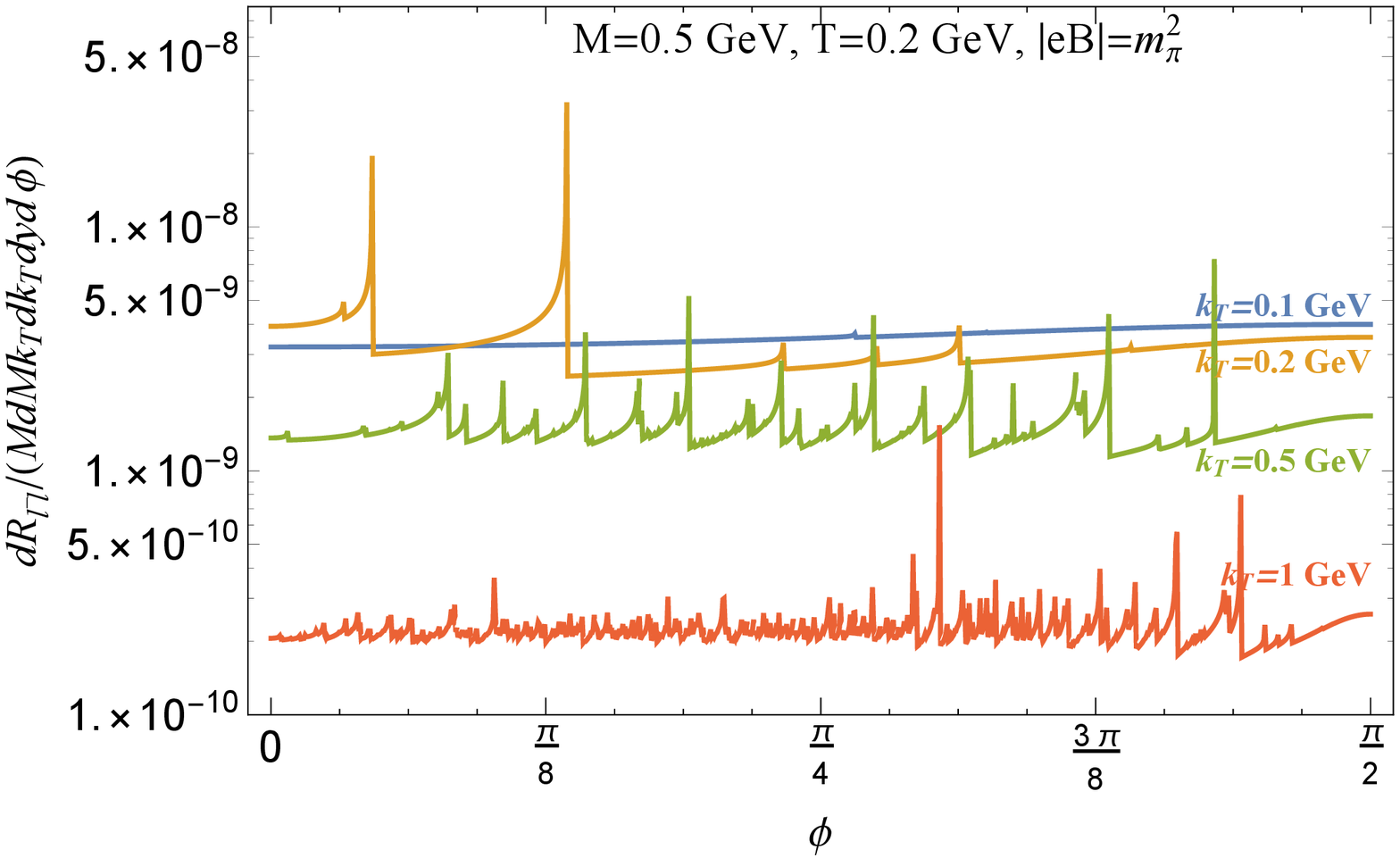}
\hspace{0.05\textwidth}
\includegraphics[width=0.46\textwidth]{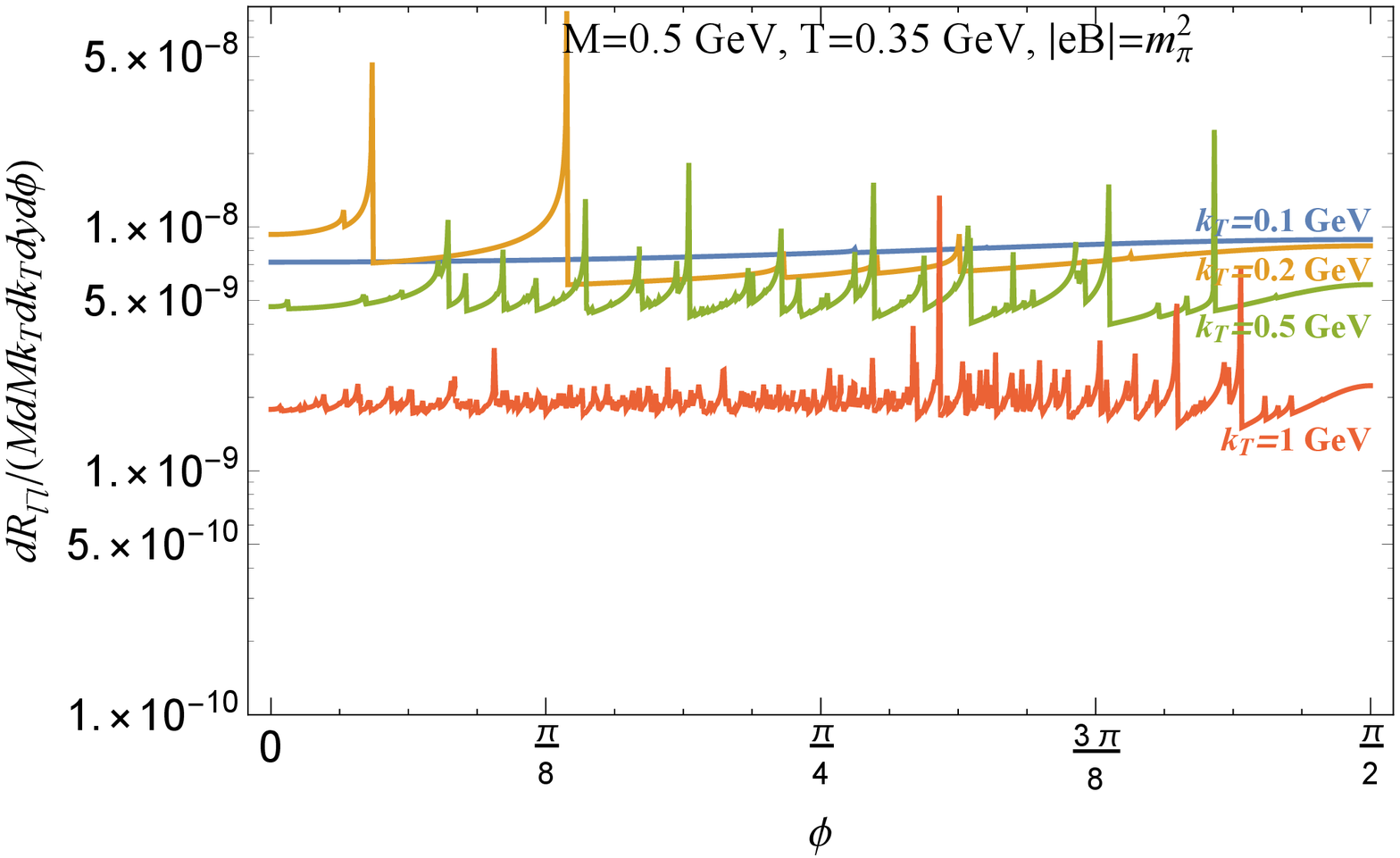}
\includegraphics[width=0.46\textwidth]{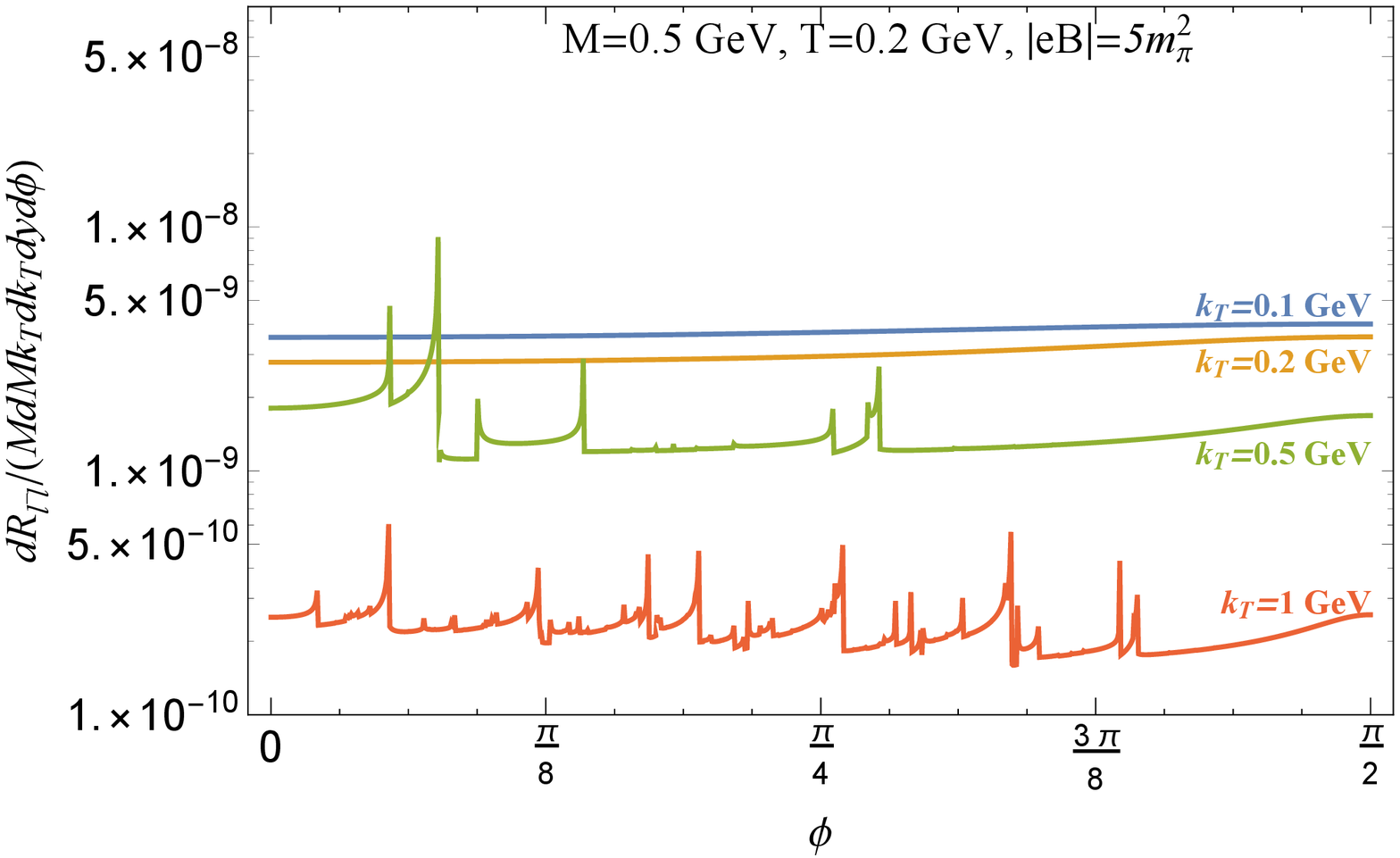}
\hspace{0.05\textwidth}
\includegraphics[width=0.46\textwidth]{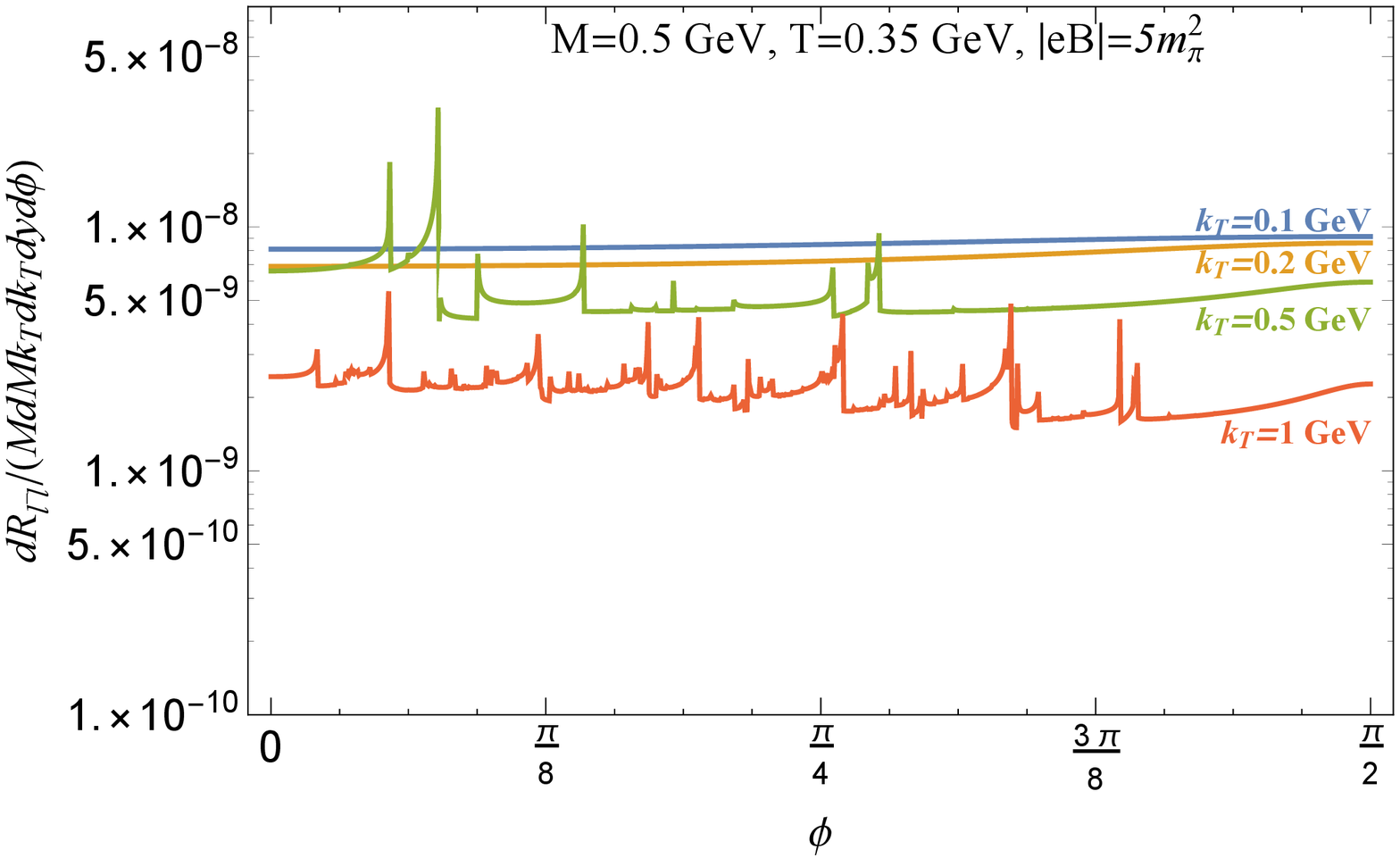}
\caption{The dilepton production rate as a function of the azimuthal angle $\phi$ for a fixed invariant mass $M = 0.5~\mbox{GeV}$, two temperatures, i.e., $T = 0.2~\mbox{GeV}$ (left panels) and $T = 0.35~\mbox{GeV}$ (right panels), and two magnetic fields, i.e., $|eB| = m_{\pi}^2$ (top panels) and $|eB| = 5m_{\pi}^2$ (bottom panels). Each panel shows the rates for four fixed transverse momenta, i.e., $k_T = 0.1~\mbox{GeV}$ (blue), $k_T = 0.2~\mbox{GeV}$ (orange), $k_T = 0.5~\mbox{GeV}$ (green), and $k_T = 1~\mbox{GeV}$ (red).}
\label{Fig:rate-vs-phi_M05}
\end{figure}

\begin{figure}[t]
\centering
\includegraphics[width=0.46\textwidth]{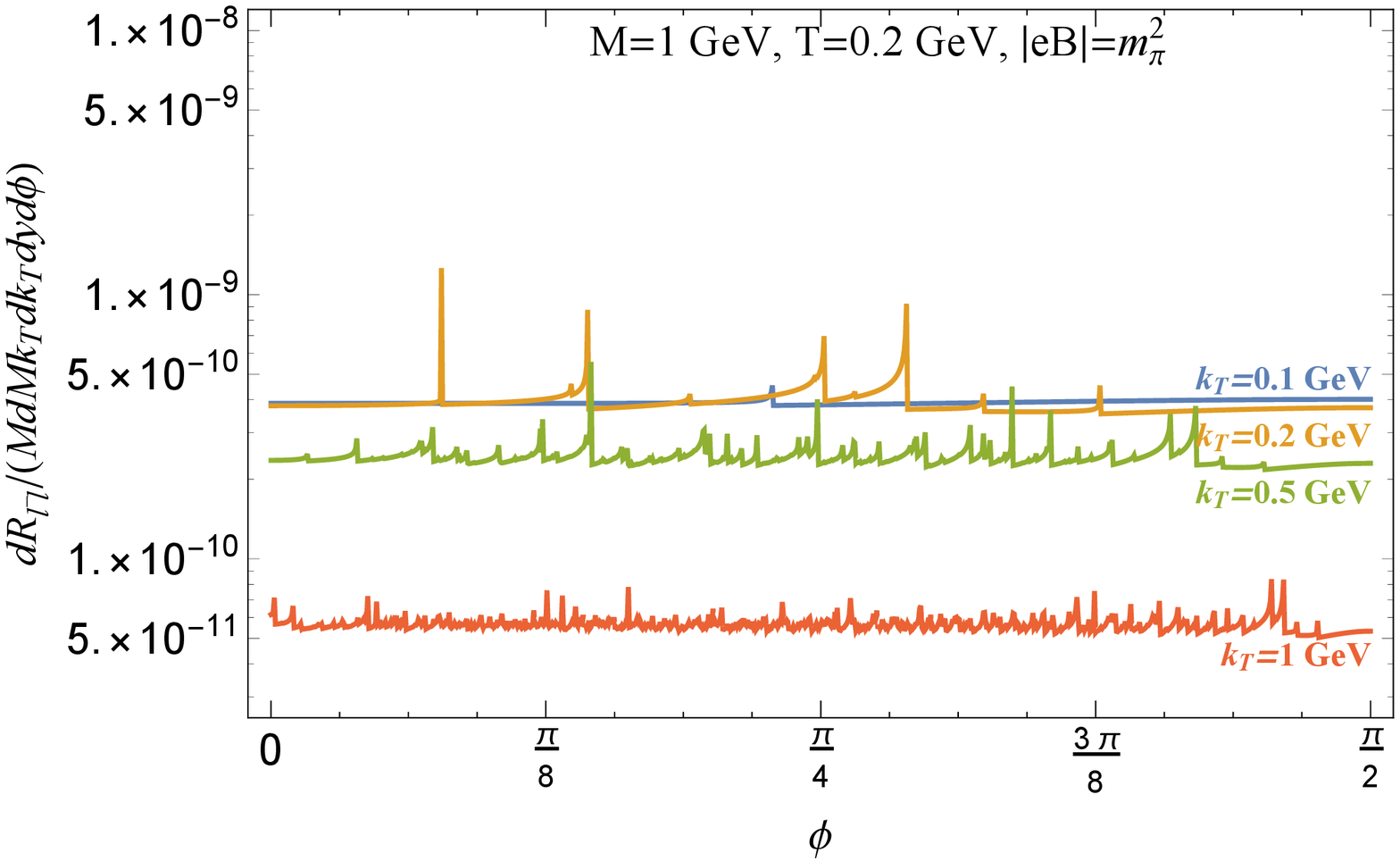}
\hspace{0.05\textwidth}
\includegraphics[width=0.46\textwidth]{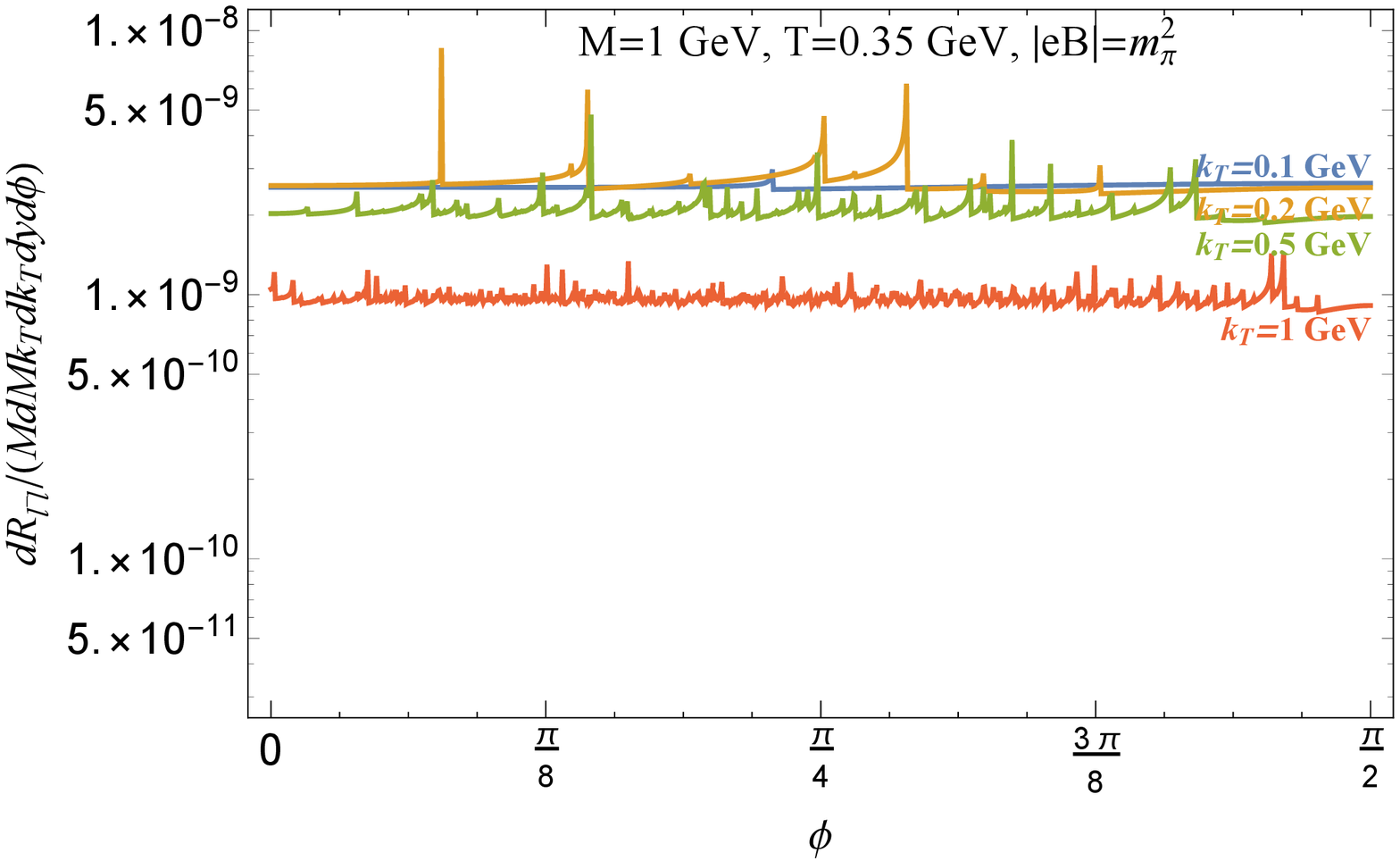}
\includegraphics[width=0.46\textwidth]{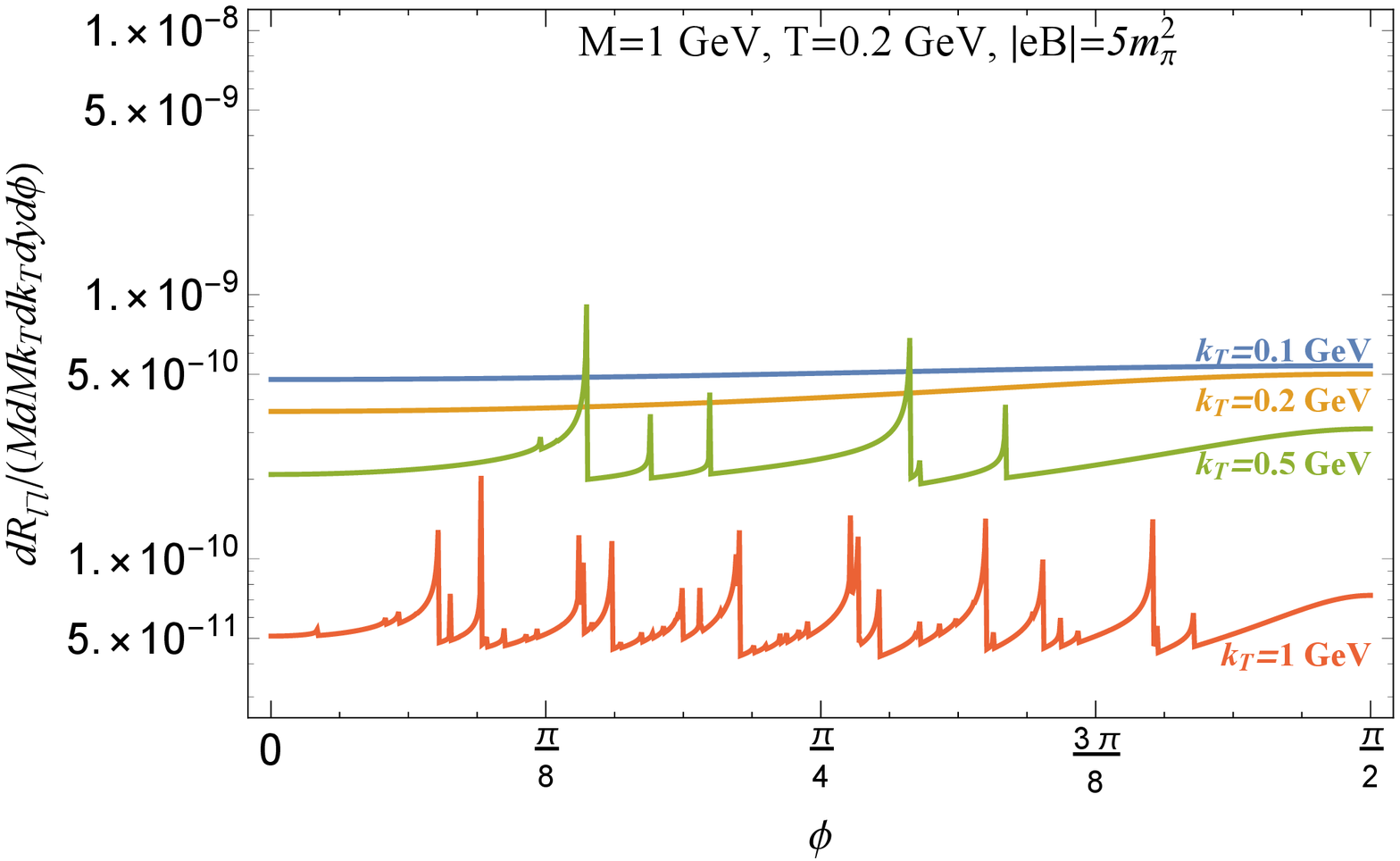}
\hspace{0.05\textwidth}
\includegraphics[width=0.46\textwidth]{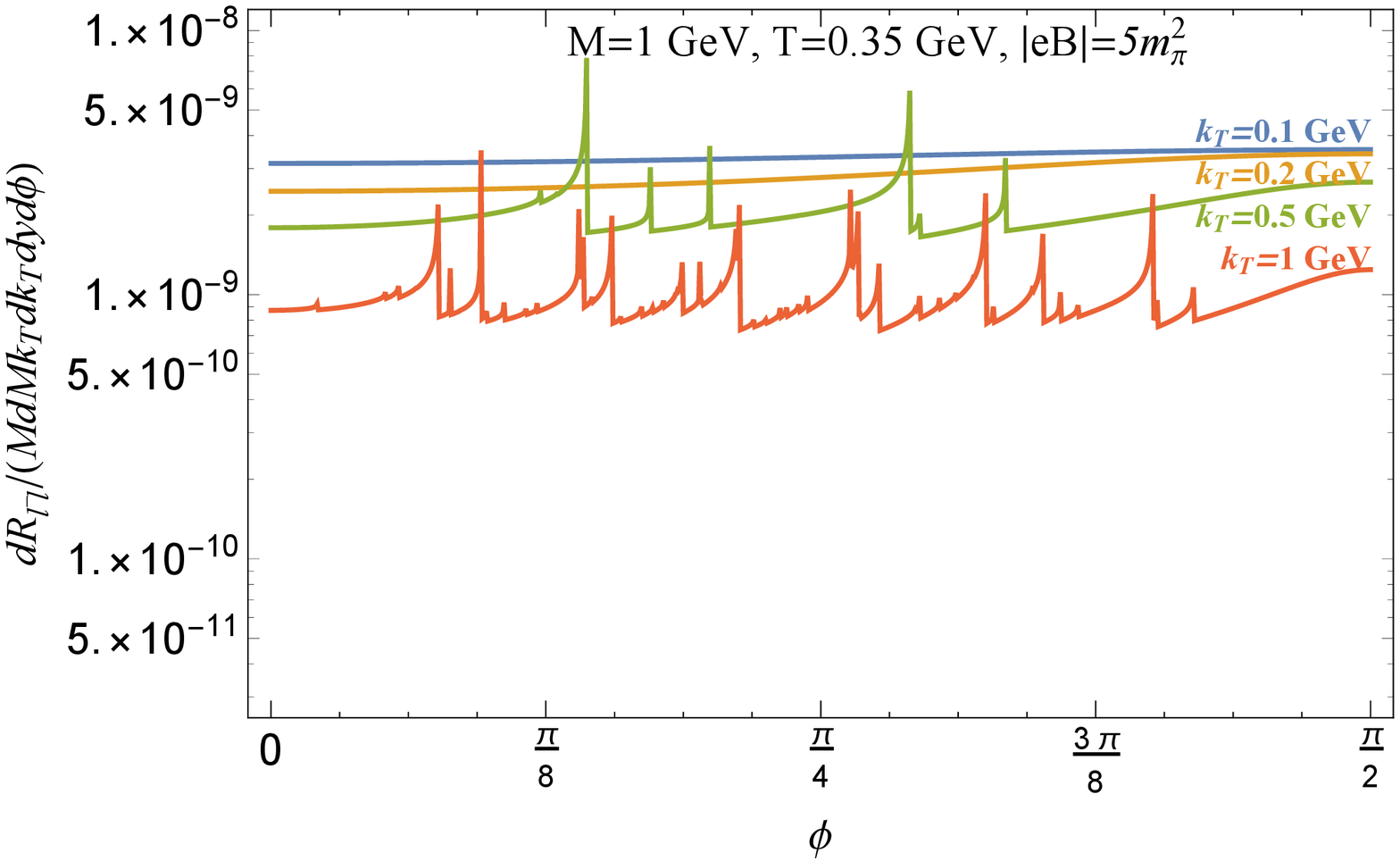}
\caption{The dilepton production rate as a function of the azimuthal angle $\phi$ for a fixed invariant mass $M = 1~\mbox{GeV}$, two temperatures, i.e., $T = 0.2~\mbox{GeV}$ (left panels) and $T = 0.35~\mbox{GeV}$ (right panels), and two magnetic fields, i.e., $|eB| = m_{\pi}^2$ (top panels) and $|eB| = 5m_{\pi}^2$ (bottom panels). Each panel shows the rates for four fixed transverse momenta, i.e., $k_T = 0.1~\mbox{GeV}$ (blue), $k_T = 0.2~\mbox{GeV}$ (orange), $k_T = 0.5~\mbox{GeV}$ (green), and $k_T = 1~\mbox{GeV}$ (red).}
\label{Fig:rate-vs-phi_M1}
\end{figure}

Because of the presence of a background magnetic field, the differential dilepton rate is expected to be anisotropic. To quantify the anisotropy it is instructive to investigate the detailed angular dependence of the rate. The corresponding representative results for several fixed values of the invariant mass, i.e., $M=0.02~\mbox{GeV}$, $M=0.5~\mbox{GeV}$, and $M=1~\mbox{GeV}$, are shown Figs.~\ref{Fig:rate-vs-phi_M002}, \ref{Fig:rate-vs-phi_M05}, and \ref{Fig:rate-vs-phi_M1}, respectively. Each figure contains a set of four panels with the rates for two values of the temperature, i.e., $T = 0.2~\mbox{GeV}$ (left panels) and $T = 0.35~\mbox{GeV}$ (right panels), and two values of the magnetic field, i.e., $|eB| = m_{\pi}^2$ (top panels) and $|eB| = 5m_{\pi}^2$ (bottom panels). Each individual panel shows the rates for several fixed transverse momenta, i.e., $k_T = 0.1~\mbox{GeV}$ (blue), $k_T = 0.2~\mbox{GeV}$ (orange), $k_T = 0.5~\mbox{GeV}$ (green), and $k_T = 1~\mbox{GeV}$ (red).

By reviewing the numerical data for the angular dependence, one can readily identify several qualitative features in the dilepton rate. In general, the rates are nonsmooth functions of the azimuthal angle $\phi$. The numerous spikes in the rate come from same Landau-level threshold effects that showed earlier in its dependence on the invariant mass. Repeating the arguments of Ref.~\cite{Wang:2021ebh}, we know that the interaction effects in the hot QGP plasma should lead to nonzero quasiparticle widths for quarks that, in turn, should smooth out the functional dependence in Figs.~\ref{Fig:rate-vs-phi_M002}, \ref{Fig:rate-vs-phi_M05}, and \ref{Fig:rate-vs-phi_M1}. Nevertheless, the dilepton rate will still have a substantial overall anisotropy for some ranges of model parameters. We discuss the corresponding quantitative measure of anisotropy in the next subsection.

\subsection{Ellipticity of dilepton emission}

One can quantify the degree of anisotropy in the dilepton production by using the conventional ellipticity parameter $v_2$. By definition, it is defined in terms of the differential rate as follows:
\begin{equation}
\label{v2}
v_{2}(M, k_T) = \frac{\int_0^{2\pi} d\phi  \cos(2\phi) \left(d R_{l\bar{l}} /d^{4} k\right)}{\int_0^{2\pi} d\phi \left(d R_{l\bar{l}} /d^{4} k\right)} .
\end{equation}
This characteristics measures the average ellipticity of dilepton rate. While it is also affected by the Landau-level threshold effects, the overall average profile should remain mostly unchanged even after the quark interactions in the plasma are accounted for \cite{Wang:2021ebh}. 

\begin{figure}[t]
\centering
\includegraphics[width=0.46\textwidth]{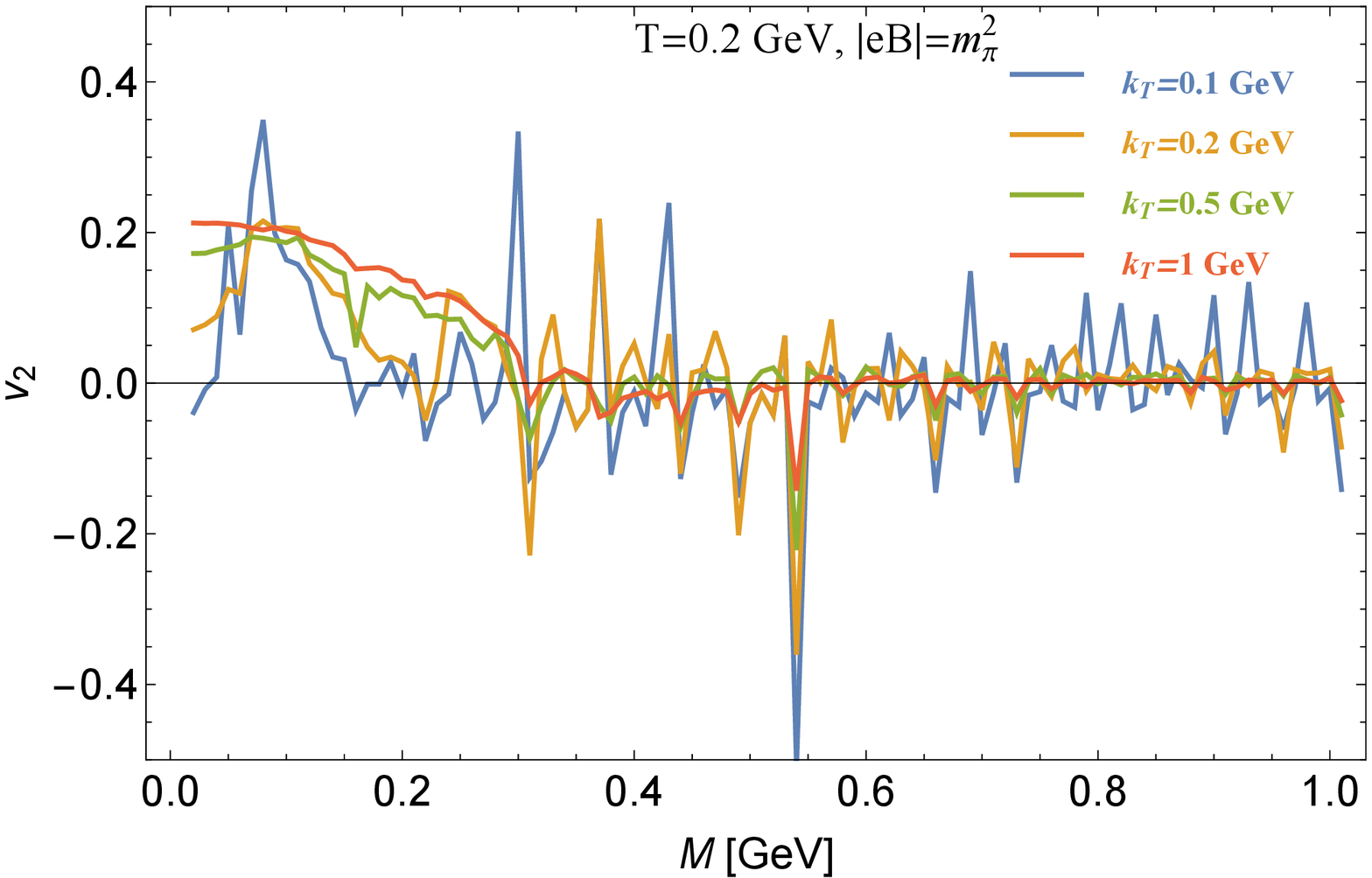}
\hspace{0.05\textwidth}
\includegraphics[width=0.46\textwidth]{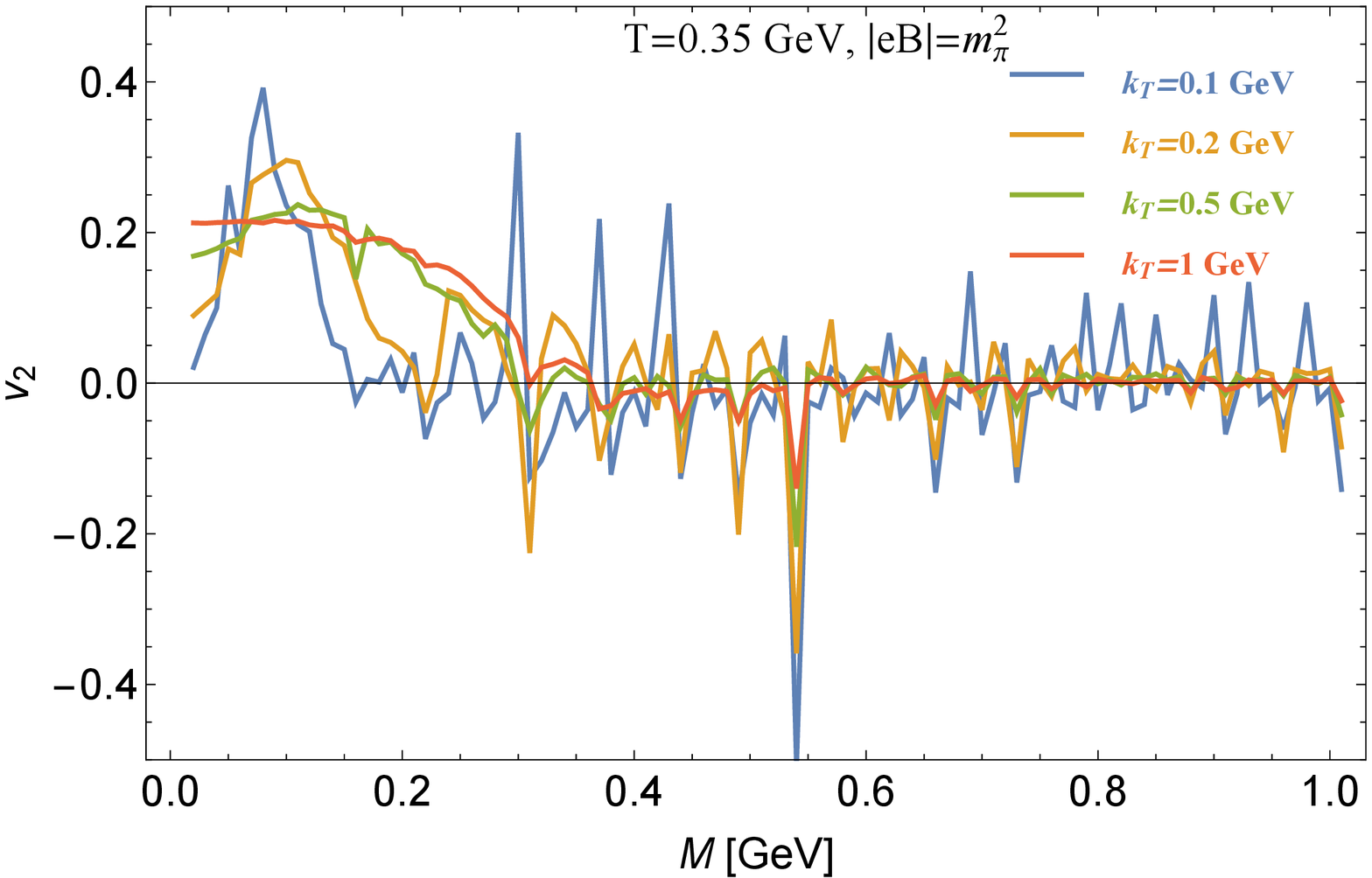}
\includegraphics[width=0.46\textwidth]{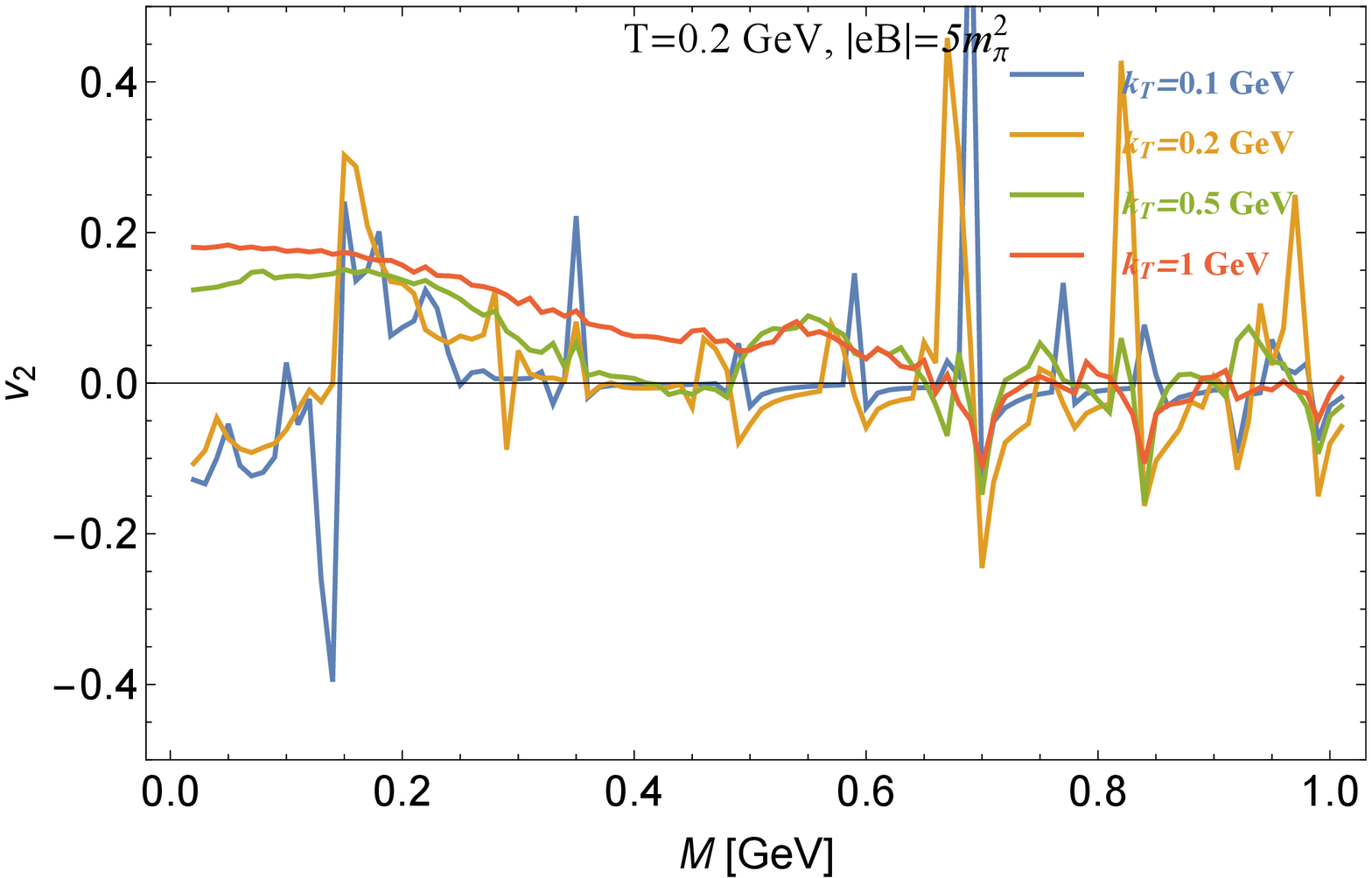}
\hspace{0.05\textwidth}
\includegraphics[width=0.46\textwidth]{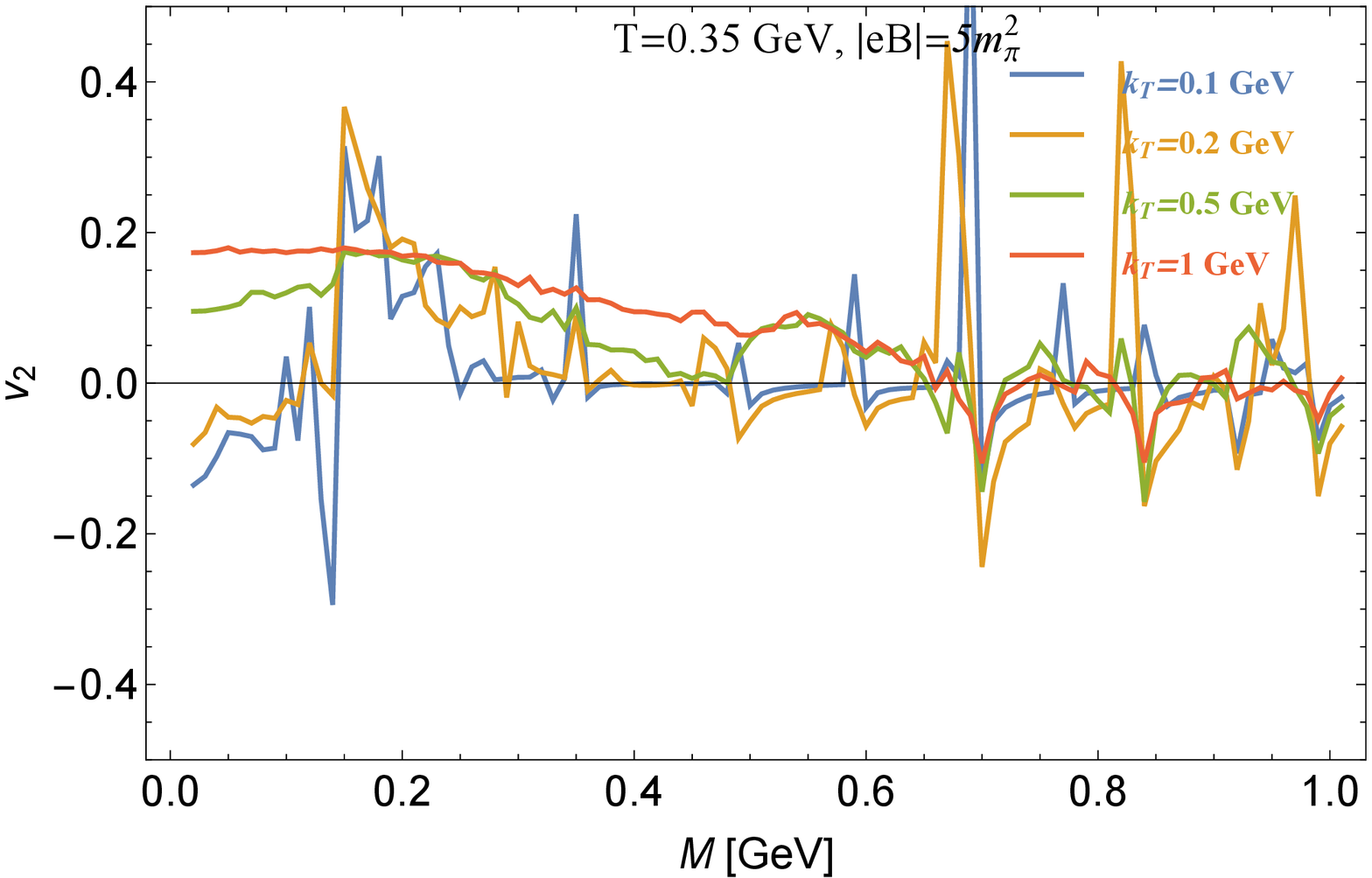}
\caption{The dilepton emission ellipticity as a function of the dilepton invariant mass $M$ for two temperatures, i.e., $T = 0.2~\mbox{GeV}$ (left panels) and $T = 0.35~\mbox{GeV}$ (right panels), and two magnetic fields, i.e., $|eB| = m_{\pi}^2$ (top panels) and $|eB| = 5m_{\pi}^2$ (bottom panels). Each panel shows the results for four fixed transverse momenta, i.e., $k_T = 0.1~\mbox{GeV}$ (blue), $k_T = 0.2~\mbox{GeV}$ (orange), $k_T = 0.5~\mbox{GeV}$ (green), and $k_T = 1~\mbox{GeV}$ (red).}
\label{Fig:v2-vs-M}
\end{figure}

The dependence of the ellipticity parameter $v_2$ on the dilepton invariant mass $M$ is shown in Fig.~\ref{Fig:v2-vs-M}. The four panels contain the results for same two temperature, i.e., $T = 0.2~\mbox{GeV}$ (two left panels) and $T = 0.35~\mbox{GeV}$ (two right panels), and two values of the magnetic field, i.e., $|eB| = m_{\pi}^2$ (two top panels) and $|eB| = 5m_{\pi}^2$ (two bottom panels). Each panel shows the ellipticities for the four different values of the transverse momenta: $k_T = 0.1~\mbox{GeV}$ (blue), $k_T = 0.2~\mbox{GeV}$ (orange), $k_T = 0.5~\mbox{GeV}$ (green), and $k_T = 1~\mbox{GeV}$ (red).

From Fig.~\ref{Fig:v2-vs-M}, in the case of small invariant masses, $M\lesssim \sqrt{|eB|}$, there is a clear tendency of $v_2$ to become positive. The latter is particularly well pronounced at large $k_T$, which is analogous to the synchrotron regime in the photon emission \cite{Wang:2021ebh}. Such a behavior is consistent with the angular dependence of the rate in Fig.~\ref{Fig:rate-vs-phi_M002}, which shows the data for the smallest invariant mass $M = 0.02~\mbox{GeV}$. The corresponding rates are clearly anisotropic when the transverse momenta take two larger values, i.e., $k_T = 0.5~\mbox{GeV}$ (green lines) or $k_T = 1~\mbox{GeV}$ (red lines). Indeed,  as one can see, the overall values are systematically larger for the azimuthal directions near $\phi \approx 0$ (in the reaction plane) and systematically smaller for $\phi \approx \pi/2$ (out of the reaction plane). Qualitatively, this implies that the differential production rate has an oblate shape ($v_2>0$). 

\begin{figure}[t]
\centering
\includegraphics[width=0.46\textwidth]{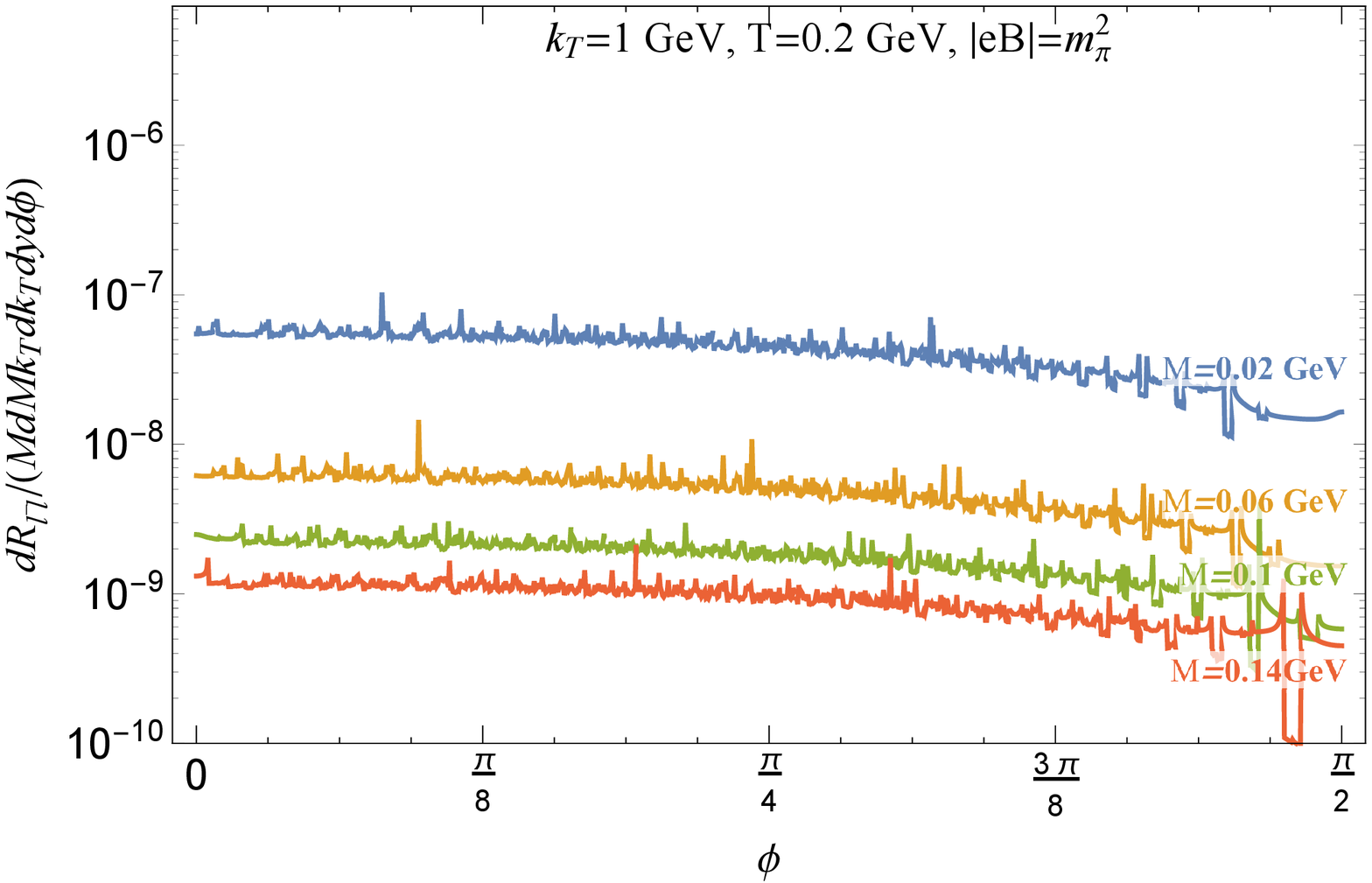}
\hspace{0.05\textwidth}
\includegraphics[width=0.46\textwidth]{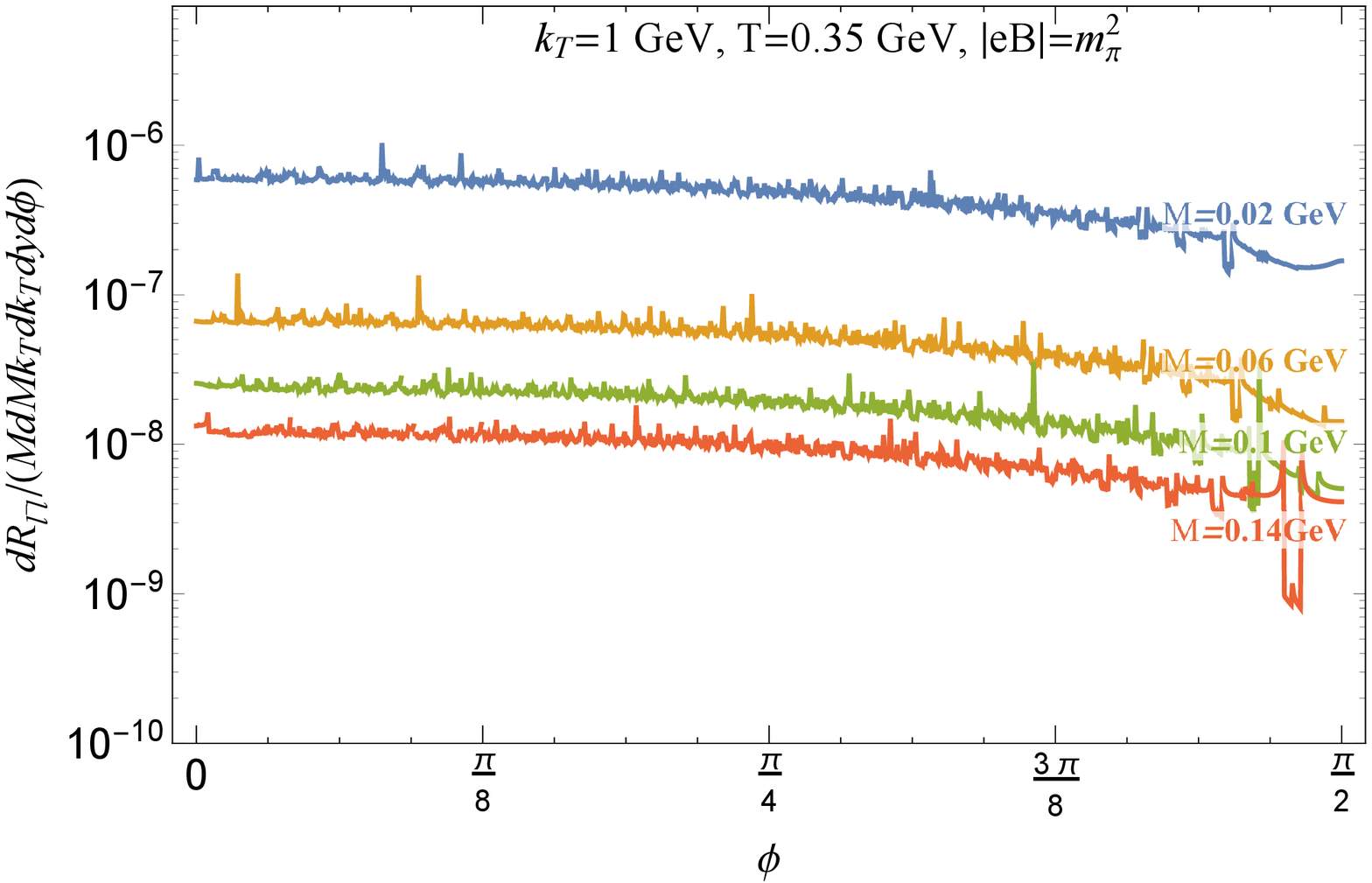}
\includegraphics[width=0.46\textwidth]{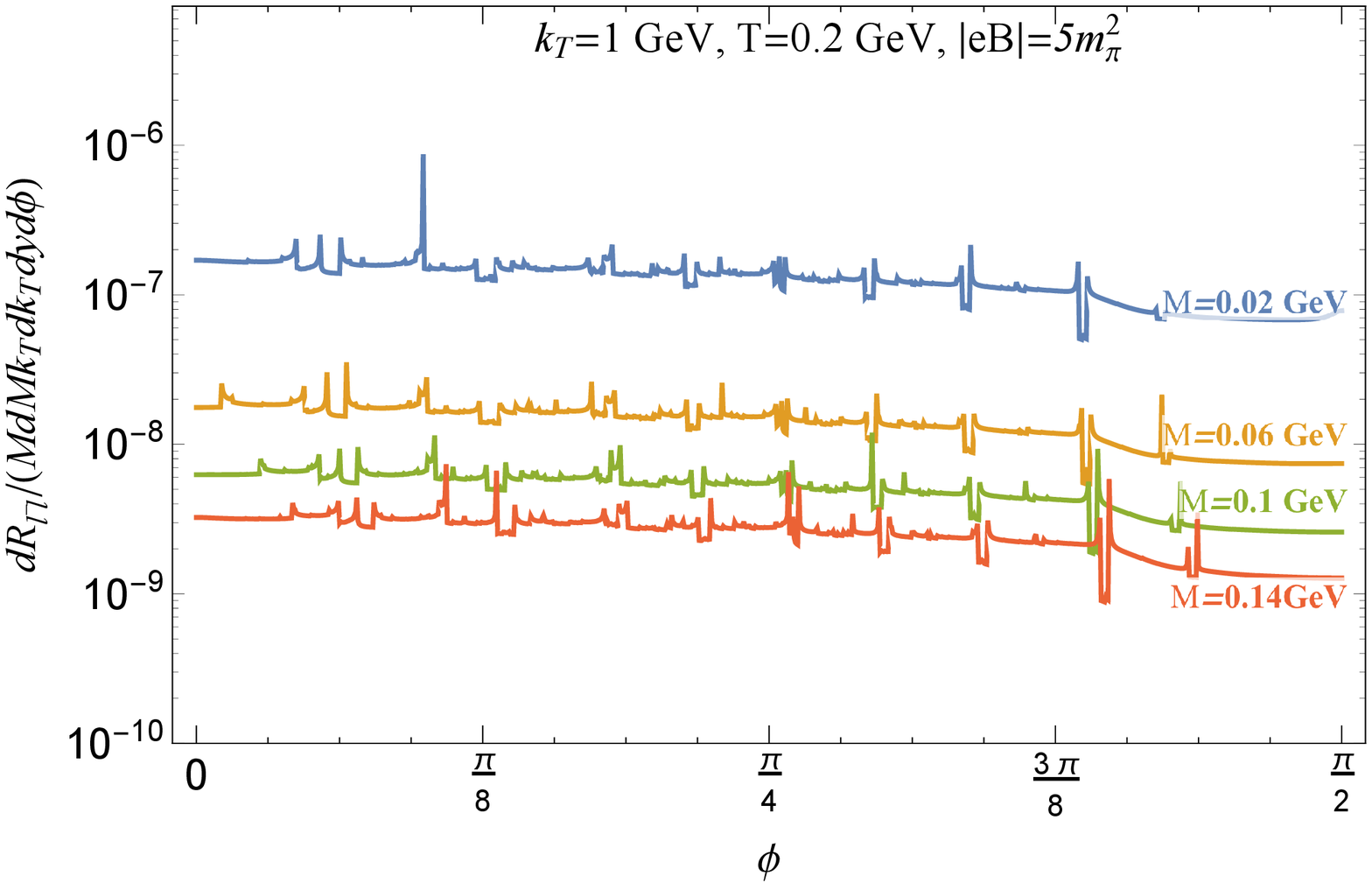}
\hspace{0.05\textwidth}
\includegraphics[width=0.46\textwidth]{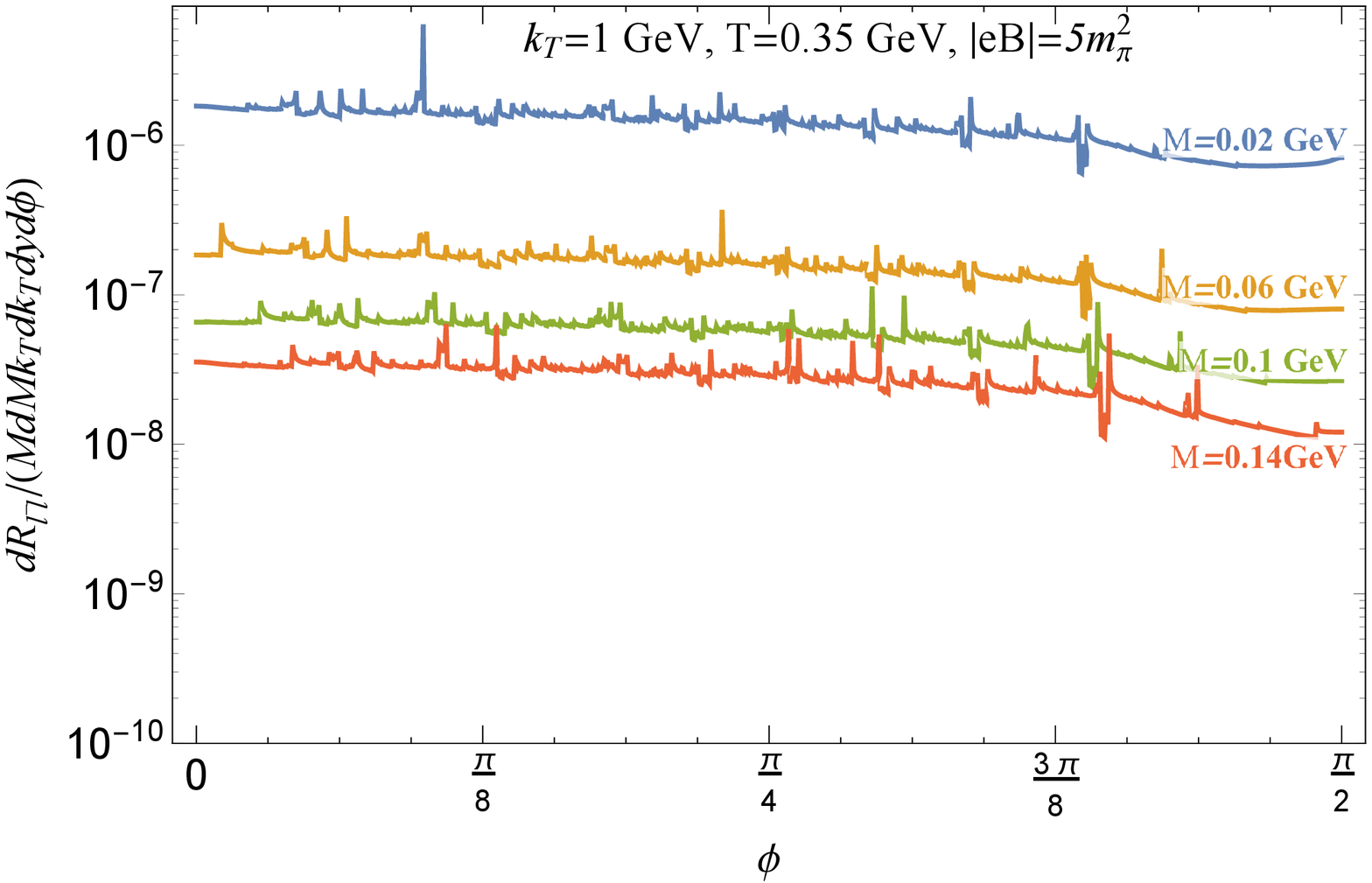}
\caption{The dilepton production rate as a function of the azimuthal angle $\phi$ for a fixed transverse momentum $k_T = 1~\mbox{GeV}$, two temperatures, i.e., $T = 0.2~\mbox{GeV}$ (left panels) and $T = 0.35~\mbox{GeV}$ (right panels), and two magnetic fields, i.e., $|eB| = m_{\pi}^2$ (top panels) and $|eB| = 5m_{\pi}^2$ (bottom panels). Each panel shows the rates for four fixed invariant masses, i.e., $M = 0.02~\mbox{GeV}$ (blue), $M = 0.06~\mbox{GeV}$ (orange), $M = 0.1~\mbox{GeV}$ (green), and $M = 0.14~\mbox{GeV}$ (red).}
\label{Fig:rate-vs-phi_kT1}
\end{figure}

The positive value of the ellipticity parameter $v_2$ for the dilepton rate at small invariant masses, $M\lesssim \sqrt{|eB|}$, and large transverse momenta can be verified by plotting the angular dependence of the rates at fixed $k_T=1~\mbox{GeV}$ and several fixed small values of $M$. The corresponding data is shown in Fig.~\ref{Fig:rate-vs-phi_kT1}. As one can see, the data confirms that the average rate is highest near $\phi \approx 0$ and smallest near $\phi \approx \pi/2$. 

The overall ellipticity is harder to discern from the data at small values of the transverse momenta, i.e., $k_T = 0.1~\mbox{GeV}$ (blue lines) and $k_T = 0.2~\mbox{GeV}$ (orange lines). For sufficiently small invariant masses, $M\lesssim \sqrt{|eB|}$, the ellipticity parameter appears to be generally nonvanishing. However, as seen from Fig.~\ref{Fig:v2-vs-M}, its sign may change from being positive at intermediate values of $M$ (i.e., $M\simeq \sqrt{|eB|}$), to being negative at sufficiently small values of $M$ (i.e., $M\ll \sqrt{|eB|}$). This is particularly clear in the case of the stronger magnetic field $|eB| = 5m_{\pi}^2$, illustrated by the two bottom panels in Fig.~\ref{Fig:v2-vs-M}.
 
It should be also noted that the average ellipticity parameter $v_2$ is consistent with zero for sufficiently large invariant masses, $M\gtrsim \sqrt{|eB|}$, see Fig.~\ref{Fig:v2-vs-M}. This conclusion is also reconfirmed by the angular dependence of the rate in Fig.~\ref{Fig:rate-vs-phi_M1}, which shows the data for the largest invariant mass $M = 1~\mbox{GeV}$ and four different values of the transverse momenta. Indeed, the corresponding rates have overall flat profiles.

\section{Discussion and conclusion}
\label{sec-4}

In this paper, we derived an explicit expression for the dilepton production rate from a hot QGP in a quantizing background magnetic field. We used the Landau-level representation for the imaginary part of the photon polarization tensor. The latter has a clear structure and allows straightforward interpretation of all underlying processes in terms of quantum transitions between the Landau-level states of quarks. 

We studied in detail the numerical dependence of the dilepton rate on the invariant mass, the transverse momentum, and the azimuthal angle measured from the reaction plane. To understand the interplay of the thermal and magnetic effects, we  calculated the rates for two representative temperatures and magnetic fields. Because of the Landau-level quantization of quark in a magnetized plasma, the functional dependence of the dilepton rate has numerous threshold spikes. They appear when the center-of-mass energy crosses any of the numerous Landau-level thresholds, leading to sudden changes of the kinematic phase space for the dilepton production. As in the case of the photon emission, the functional dependence is expected to smooth out when the strong quark interactions are taken into account \cite{Wang:2021ebh}. For the QGP produced in heavy-ion collisions, the time evolution of the magnetic field should provide additional smoothing. Thus, in this study, we did not concentrate too much on the nonsmooth behavior associated with the Landau-level thresholds.

Unlike the zero-field Born rate, the dilepton rate for a magnetized QGP is generally anisotropic. The anisotropy is pronounced the most in the regime of small values of the dilepton invariant mass, i.e., $M\lesssim \sqrt{|eB|}$. When the transverse momenta are large, the ellipticity parameter $v_2$ tends to be positive and large. It resembles a similar anisotropy in the quantum synchrotron regime of photon emission at large $k_T$ \cite{Wang:2021ebh}. The situation is more subtle at small transverse momenta. In this case, the ellipticity parameter $v_2$ may still remain nonzero in general. However, its sign gradually changes from being positive at intermediate values of $M$ (i.e., $M\simeq \sqrt{|eB|}$) to being negative at small $M$ (i.e., $M\ll \sqrt{|eB|}$). Under optimal conditions, we find that the magnitude of $v_2$ could be as large as $0.2$. If such large values are measured in an experiment, they will most likely indicate the presence of a nonzero magnetic field in the QGP plasma. 

Besides inducing a nontrivial angular dependence, a strong magnetic field (i.e., $|eB|\gtrsim m_\pi^2$) substantially modifies the overall integrated dilepton rate. In particular, the background field strongly enhances the rate at small values of the dilepton invariant mass ($M\lesssim\sqrt{|eB|}$). At large values of the invariant mass ($M\gtrsim\sqrt{|eB|}$), of course, the role of the magnetic fields decreases, and the results gradually approach the isotropic zero-field Born rate. 

We argue that the significant enhancement of the integrated dilepton rate at small invariant masses is a unique signature of a nonzero background magnetic field. Thus, measuring the corresponding rate at $M\lesssim 0.2~\mbox{GeV}$, for example, could provide sufficient information to confirm or rule out the fields of order $|eB|\simeq m_\pi^2$ in relativistic heavy-ion collisions. Optimistically, the excess of measured emission over the Born rate can also constrain the value of the average field strength. To implement this proposal in practice, one will require to perform additional studies and modeling. It will be critical, for example, to investigate the emission at nonzero rapidity, to model the dependence on centrality, to estimate the background effects, and much more. 
 
\acknowledgements
The work of X.W. was supported by the start-up funding No.~4111190010 of Jiangsu University and NSFC under Grant No.~11735007. 
The work of I.A.S. was supported by the U.S. National Science Foundation under Grants No.~PHY-1713950 and No.~PHY-2209470. 

\appendix

\section{Optical conductivity}
\label{AppA}

Here we present explicit expressions for the longitudinal and transverse components of the optical conductivity of a strongly magnetized QGP. For a generic plasma, both components of the conductivity were calculated in Ref.~\cite{Wang:2021ebh}. In the case of the QGP with two lightest quark flavors, the corresponding  expressions read 
\begin{eqnarray}
\sigma_{\parallel}(\Omega) &=& \sum_{f=u,d}\frac{4\alpha N_{c}  q_f^2 }{  \Omega^2  \ell_f^2} \tanh\left(\frac{|\Omega|}{4T}\right) \sum_{n=0}^{n^{\parallel}_{\rm max}} \left(2-\delta_{n,0}\right)
\frac{M_{n,f}^2 \theta\left( \Omega^2  - 4 M_{n,f}^2\right)  }{\sqrt{\Omega^2-4 M_{n,f}^2} }  ,
\label{sigma-parallel}\\
%%%%%%%%%%%%
\sigma_{\perp}(\Omega)  &=& \sum_{f=u,d}\frac{2\alpha N_{c}  q_f^2  \sinh\left(\frac{\Omega}{2T}\right) }{\Omega \ell_{f}^4\left[\cosh\left(\frac{\Omega}{2T}\right) +\cosh\left(\frac{|e_fB|}{T\Omega}\right) \right]} \sum_{n=1}^{n^{\perp}_{\rm max}} 
\frac{ \left[  2 (2n-1) -\Omega^2 \ell_{f}^2\right] 
\theta\left[ (M_{n,f}-M_{n-1,f})^2 - \Omega^2 \right] }{
\sqrt{ \left[ (M_{n,f}-M_{n-1,f})^2 - \Omega^2 \right] \left[ (M_{n,f}+M_{n-1,f})^2 - \Omega^2 \right]} } \nonumber\\
%%%%%%%%%%%%
&+&  \sum_{f=u,d}\frac{2\alpha N_{c}  q_f^2 \sinh\left(\frac{\Omega}{2T}\right) }{\Omega \ell_{f}^4 \left[\cosh\left(\frac{\Omega}{2T}\right) +\cosh\left(\frac{|e_fB|}{T\Omega}\right) \right]} \sum_{n=1}^{n^{\perp}_{\rm max}} 
\frac{\left[ \Omega^2 \ell_{f}^2 -  2(2n-1) \right]\theta\left[  \Omega^2-(M_{n,f}+M_{n-1,f})^2  \right] }{ \sqrt{\left[  \Omega^2-(M_{n,f}-M_{n-1,f})^2  \right] \left[  \Omega^2-(M_{n,f}+M_{n-1,f})^2  \right] } }  ,
\label{sigma-perp}
\end{eqnarray}
where $N_c=3$ is the number of colors and $M_{n,f}=\sqrt{2n|e_fB|+m^2}$. Notice that the Landau-level sums in the longitudinal and transverse conductivities are finite, terminating at $n^{\parallel}_{\rm max} =\left[(\Omega^2-4m^2)/(4|e_fB|)\right]$ and $n^{\perp}_{\rm max} =\left[((2|e_fB|+\Omega^2)^2-4m^2\Omega^2)/(8|e_fB|\Omega^2)\right]$, respectively. 

In the limit of the vanishing magnetic field, the sum over Landau levels turns into an integral over the continuous variable $u =2n|e_f B|$. Therefore, in the limit $B\to 0$, one obtains
\begin{eqnarray}
\left. \sigma_{\parallel}(\Omega) \right|_{B\to 0} &=& \sum_{f=u,d}\frac{4\alpha N_{c}  q_f^2 }{  \Omega^2 } \tanh\left(\frac{|\Omega|}{4T}\right) \int_{0}^{\infty}  du
\frac{\left(u+m^2\right) \theta\left( \Omega^2  - 4 (u+m^2)\right)  }{\sqrt{\Omega^2-4 (u+m^2)} }  ,
\label{sigma-parallel-B0}\\
%%%%%%%%%%%%
\left.\sigma_{\perp}(\Omega) \right|_{B\to 0}   &=& \sum_{f=u,d}\frac{ \alpha N_{c}  q_f^2 \sinh\left(\frac{|\Omega|}{2T}\right) }{\Omega^2 \left[\cosh\left(\frac{\Omega}{2T}\right) +1 \right]} \int_{0}^{\infty} du
\frac{\left( \Omega^2  -  2u \right)\theta\left[  \Omega^2-4(u+m^2)  \right] }{  \sqrt{ \Omega^2-4(u+m^2) } }  .
\label{sigma-perp-B0}
\end{eqnarray}
In this limit, only the quark-antiquark processes contribute to the conductivity. After performing the integrals, we find that both longitudinal and transverse components of the optical conductivity reduce to the same expression:
\begin{eqnarray}
\left. \sigma_\parallel(\Omega)\right|_{B\to 0} = \left.\sigma_\perp(\Omega)\right|_{B\to 0}  &=&\frac{ \alpha N_{c} (q_u^2+q_d^2)}{3} \Omega \,\tanh\left(\frac{\Omega}{4T}\right) \sqrt{1  - \frac{4 m^2}{\Omega^2}}\left( 1  +\frac{2 m^2}{\Omega^2 }\right)\theta\left( \Omega^2  - 4 m^2\right)
\nonumber\\
&\simeq &\frac{ \alpha N_{c} (q_u^2+q_d^2)}{3} \Omega \,\tanh\left(\frac{\Omega}{4T}\right), \quad\mbox{for} \quad m\to 0.
\end{eqnarray} 

Interestingly, one can also derive a simple analytical asymptote for $\sigma_{\perp}(\Omega)$ in the limit $\Omega\ll \sqrt{|eB|}$. In this regime, the transverse conductivity is dominated by the Landau levels with high indices $n\lesssim n^{\perp}_{\rm max} \simeq  |e_fB|/(2\Omega^2)$. (Note that the value of $n^{\perp}_{\rm max}$ is consistent with the cutoff estimate $n_{\rm max}\simeq |eB|/\Omega^{2}$, obtained in the main text from the quantization of low-lying Landau levels.) Thus, by approximating the sum in Eq.~(\ref{sigma-perp}) with an integral over $u=2n(\Omega\ell_{f})^2$, we derive
\begin{equation}
\sigma_{\perp}(\Omega)  \simeq  \sum_{f=u,d}\frac{\alpha N_{c}  q_f^2 \sinh\left(\frac{\Omega}{2T}\right) }{\Omega^5 \ell_{f}^6\left[\cosh\left(\frac{\Omega}{2T}\right) +\cosh\left(\frac{|e_fB|}{T\Omega}\right) \right]} \int_{0}^{1} \frac{u  du}{\sqrt{1-u}} 
=\sum_{f=u,d}\frac{4\alpha N_{c}  q_f^2 |e_fB|^3\sinh\left(\frac{\Omega}{2T}\right) }{3\Omega^5 \left[\cosh\left(\frac{\Omega}{2T}\right) +\cosh\left(\frac{|e_fB|}{T\Omega}\right) \right]}   .
\label{sigma-perp-Omega-to-0}
\end{equation} 
In the same limit $\Omega\ll \sqrt{|eB|}$, the longitudinal component $\sigma_{\parallel}(\Omega)$ is negligible. It is suppressed by a factor of the order of $\left(m \Omega/|eB|\right)^2$ compared to $\sigma_{\perp}(\Omega)$. 

Let us emphasize in passing that  Eq.~(\ref{sigma-perp-Omega-to-0}) represents the strong-field limit but not the lowest Landau level approximation. Actually, it is the opposite of the lowest Landau level approximation since many Landau levels, with indices up to $n_{\rm max} \simeq |e_fB|/(2\Omega^2)$, contribute to the result.

By substituting the above results for optical conductivity into Eq.~(\ref{rate_conductivity}), one can derive an analytical expression for the dilepton rate at $k_T=0$ and $M\ll \sqrt{|eB|}$, see Eq.~(\ref{rate_conductivity_kT0-small-M}) in the main text.

\end{document}